\documentclass[longauth]{aa} 

\usepackage{graphicx}
\usepackage{txfonts}
\usepackage[utf8]{inputenc}
\usepackage{array}
\usepackage{float}
\usepackage{subfig}
\usepackage{multirow}
\usepackage[]{natbib}
\usepackage{amsmath}
\usepackage{epstopdf} 
\usepackage[colorlinks=true,linkcolor=blue,citecolor=blue]{hyperref}
\begin{document} 

   \title{VLT/SPHERE astrometric confirmation and orbital analysis of the brown dwarf companion HR~2562~B\thanks{Based on observations collected at the European Organisation for Astronomical Research in the Southern Hemisphere under ESO programme 198.C-0209.}}

   \author{A.-L. Maire\inst{1}, L. Rodet\inst{2}, C. Lazzoni\inst{3,4}, A. Boccaletti\inst{5}, W. Brandner\inst{1}, R. Galicher\inst{5}, F. Cantalloube\inst{1}, D. Mesa\inst{6,3}, H. Klahr\inst{1}, H. Beust\inst{2}, G. Chauvin\inst{2,7}, S. Desidera\inst{3}, M. Janson\inst{8,1}, M. Keppler\inst{1}, J. Olofsson\inst{9,10,1}, J.-C. Augereau\inst{2}, S. Daemgen\inst{11}, T. Henning\inst{1}, P. Th\'ebault\inst{5}, M. Bonnefoy\inst{2}, M. Feldt\inst{1}, R. Gratton\inst{3}, A.-M. Lagrange\inst{2}, M. Langlois\inst{12,13}, M.~R. Meyer\inst{14,11}, A. Vigan\inst{13}, V. D'Orazi\inst{3}, J. Hagelberg\inst{2}, H. Le Coroller\inst{13}, R. Ligi\inst{15,13}, D. Rouan\inst{5}, M. Samland\inst{1}, T. Schmidt\inst{5}, S. Udry\inst{16}, A. Zurlo\inst{17,18,13,3}, L. Abe\inst{19}, M. Carle\inst{13}, A. Delboulb\'e\inst{2}, P. Feautrier\inst{2}, Y. Magnard\inst{2}, D. Maurel\inst{2}, T. Moulin\inst{2}, A. Pavlov\inst{1}, D. Perret\inst{5}, C. Petit\inst{20}, J.~R. Ramos\inst{1}, F. Rigal\inst{21}, A. Roux\inst{2}, and L. Weber\inst{16}}

   \institute{Max-Planck-Institut f\"ur Astronomie, K\"onigstuhl 17, D-69117 Heidelberg, Germany \\
          \email{maire@mpia.de}
         \and  
         Univ. Grenoble Alpes, CNRS, IPAG, F-38000 Grenoble, France
         \and
         INAF -- Osservatorio Astronomico di Padova, Vicolo dell'Osservatorio 5, I-35122 Padova, Italy
         \and
         Dipartimento  di  Fisica  e  Astronomia  ``G.  Galilei'',  Universit\`a di Padova, Via Marzolo, 8, 35121 Padova, Italy
         \and       
         LESIA, Observatoire de Paris, PSL Research University, CNRS, Universit\'e Paris Diderot, Sorbonne Paris Cit\'e, UPMC Paris 6, Sorbonne Universit\'e, 5 place J. Janssen, F-92195 Meudon, France
         \and
         INCT, Universidad de Atacama, calle Copayapu 485, Copiap\'{o}, Atacama, Chile
         \and
         Unidad Mixta Internacional Franco-Chilena de Astronom\'ia CNRS/INSU UMI 3386 and Departamento de Astronom\'ia, Universidad de Chile, Casilla 36-D, Santiago, Chile
         \and
         Department of Astronomy, Stockholm University, AlbaNova University Center, 106 91 Stockholm, Sweden
         \and
         Instituto de F\'isica y Astronom\'ia, Facultad de Ciencias, Universidad de Valpara\'iso, Av. Gran Breta\~na 1111, Playa Ancha, Valpara\'iso, Chile
         \and
         N\'ucleo Milenio Formaci\'on Planetaria - NPF, Universidad de Valpara\'iso, Av. Gran Breta\~na 1111, Valpara\'iso, Chile
         \and
         ETH Zurich, Institute for Particle Physics and Astrophysics, Wolfgang-Pauli-Strasse 27, 8093 Zurich, Switzerland
         \and
         CRAL, UMR 5574, CNRS/ENS-Lyon/Universit\'e Lyon 1, 9 av. Ch. Andr\'e, F-69561 Saint-Genis-Laval, France
         \and
         Aix Marseille Univ., CNRS, LAM, Laboratoire d'Astrophysique de Marseille, Marseille, France
         \and
         Department of Astronomy, University of Michigan, 1085 S. University  Ave, Ann Arbor, MI 48109-1107, USA
         \and
         INAF -- Osservatorio Astronomico di Brera, via Bianchi 46, I-23807 Merate (LC), Italy   
         \and
         Geneva Observatory, University of Geneva, Chemin des Maillettes 51, 1290 Versoix, Switzerland 
         \and
         N\'ucleo de Astronom\'ia, Facultad de Ingenier\'ia, Universidad Diego Portales, Av. Ejercito 441, Santiago, Chile
         \and
         Millennium Nucleus ``Protoplanetary Disk'', Departamento de Astronom\'ia, Universidad de Chile, Casilla 36-D, Santiago, Chile
         \and
         Universit\'e C\^ote d'Azur, OCA, CNRS, Lagrange, France
         \and
         DOTA, ONERA, Universit\'e Paris Saclay, F-91123 Palaiseau, France
         \and
         Anton Pannekoek Institute for Astronomy, Science Park 904, NL-1098 XH Amsterdam, The Netherlands   
             }

   \date{Received 15 December 2017/ Accepted 9 April 2018}

 
  \abstract
   {A low-mass brown dwarf has recently been imaged around HR~2562 (HD~50571), a star hosting a debris disk resolved in the far infrared. Interestingly, the companion location is compatible with an orbit coplanar with the disk and interior to the debris belt. This feature makes the system a valuable laboratory to analyze the formation of substellar companions in a circumstellar disk and potential disk-companion dynamical interactions.}
   {We aim to further characterize the orbital motion of HR~2562~B and its interactions with the host star debris disk.}
   {We performed a monitoring of the system over $\sim$10 months in 2016 and 2017 with the VLT/SPHERE exoplanet imager.}
  {We confirm that the companion is comoving with the star and detect for the first time an orbital motion at high significance, with a current orbital motion projected in the plane of the sky of 25~mas ($\sim$0.85~au) per year. No orbital curvature is seen in the measurements. An orbital fit of the SPHERE and literature astrometry of the companion without priors on the orbital plane clearly indicates that its orbit is (quasi-)coplanar with the disk. To further constrain the other orbital parameters, we used empirical laws for a companion chaotic zone validated by N-body simulations to test the orbital solutions that are compatible with the estimated disk cavity size. Non-zero eccentricities ($>$0.15) are allowed for orbital periods shorter than 100~yr, while only moderate eccentricities up to $\sim$0.3 for orbital periods longer than 200~yr are compatible with the disk observations. A comparison of synthetic \textit{Herschel} images to the real data does not allow us to constrain the upper eccentricity of the companion.}
   {}


\authorrunning{A.-L. Maire et al.}
\titlerunning{}

   \maketitle
%
\section{Introduction}

\begin{table*}[t]
\caption{Observing log.}
\label{tab:obs}
\begin{center}
\begin{tabular}{l c c c c c c c c}
\hline\hline
UT date & $\epsilon$ ($''$) & $\tau_0$ (ms) & AM start/end & Mode & Bands & DIT\,(s)\,$\times$\,Nfr & $\Delta$PA ($^{\circ}$) & SR \\
\hline
2016/12/12 & 1.6--2.2 & 2 & 1.23--1.25 & IRDIFS & $YJ$+$H$ & 16(64)$\times$256(64) & 28.1 & 0.30--0.53 \\
2017/02/07 & 0.4--0.9 & 4--8 & 1.24--1.26 & IRDIFS & $YJ$+$H$ & 16(64)$\times$320(80) & 34.9 & 0.72--0.92 \\
2017/09/29 & 0.5--1.0 & 2--4 & 1.39--1.26 & IRDIFS\_EXT & $YJH$+$K1K2$ & 48(64)$\times$100(75) & 27.6 & 0.69--0.87 \\
\hline
\end{tabular}
\end{center}
\tablefoot{The columns provide the observing date, the seeing {measured by the differential image motion monitor (DIMM) at 0.5~$\muup$m, the associated coherence time}, the airmass (AM) at the beginning and the end of the sequence, the observing mode, the spectral bands, the {detector integration time (DIT)} multiplied by the number of frames in the sequence {(Nfr)}, the field of view rotation, and the Strehl ratio measured by the {adaptive optics} system at 1.6~$\muup$m. For the DIT$\times$Nfr column, the numbers in parentheses are for the IFS data.}
\end{table*} 

HR~2562 (HD~50571, HIP~32775) is a nearby F5V star of mass 1.3~$M_\odot$ \citep{Gray2006, Casagrande2011} with high proper motion \citep[$d$~=~33.64$\pm$0.45~pc, $\mu_{\mathrm{\alpha}}$~=~4.872$\pm$0.040~mas/yr, $\mu_{\mathrm{\delta}}$~=~108.568$\pm$0.040~mas/yr,][]{GaiaCollaboration2016} known to host an extended debris disk of outer radius 187\,$\pm$\,20~au with a fractional luminosity of the infrared excess $L_{\mathrm{disk}}/L_{\star}$\,=\,(1.0$\pm$0.3)\,$\times$\,10$^{-4}$ and a large inner hole of radius $\sim$18--70~au \citep{Moor2006, Moor2011, Moor2015}, and a late-L brown dwarf companion \citep{Konopacky2016b}.

Modelings of the stellar spectral energy distribution (SED) show evidence for a single cold ($\sim$40--70~K) outer component \citep{Moor2011, Pawellek2014, Moor2015}. Pure SED fittings give a cold disk average radius of 58--71~au \citep{Moor2011, Moor2015}, whereas \textit{Herschel}/PACS image fittings point towards a larger average radius, 104--138~au \citep{Pawellek2014, Moor2015}. \citet{Kral2017} estimated this parameter to be 181~au from an SED fit combined with the blackbody radius correction proposed by \citet{Pawellek2015} assuming an equal mixture of ices and astrosilicates. The inclination and the position angle of the disk were estimated to be 78.0$\pm$6.3$^{\circ}$ and 120.1$\pm$3.2$^{\circ}$ by \citet{Moor2015} from the fit of a geometrical disk model to \textit{Herschel}/PACS images. \citet{Pawellek2014} estimated an index for the size distribution of the dust grains of 4.01$\pm$0.49, which is consistent with predictions from collisional cascade models \citep[e.g.,][]{Dohnanyi1969, Thebault2008, Krivov2013}.

The age estimate of the system is quite uncertain with a range of 300--900~Myr \citep{Konopacky2016b}, translating into a mass range for the brown dwarf companion of 15--45~$M_{\rm{J}}$. Thanks to the large stellar proper motion and the high astrometric accuracy of the GPI instrument, \citet{Konopacky2016b} were able to confirm that the companion is physically bound to HR~2562 using observations taken one month apart. Although the limited orbital coverage prevent them from performing an orbital analysis of the companion, they noted that its location is compatible with an orbital plane coplanar with the debris disk with a projected separation ($\sim$20~au) interior to the debris belt. Together with the HD~206893 system \citep{Milli2017a}, HR~2562 therefore represents a valuable laboratory for studying the formation of substellar companions in a circumstellar disk in a higher-mass regime with respect to other known debris disk systems hosting planetary mass companions \citep[e.g., HR~8799, $\beta$~Pic, HD~95086, 51~Eri,][]{Marois2008c, Marois2010b, Lagrange2010b, Rameau2013c, Macintosh2015}. Recently, {\citet{Mesa2018} reassessed the stellar properties using the isochrones of \citet{Bressan2012} and a Bayesian determination approach \citep[see details in][]{Desidera2015}} and found a mass value similar to previous estimates (1.368$\pm$0.018~$M_\odot$) but a slightly younger age range of 200--750~Myr.

We present in this paper high-contrast images of HR~2562 obtained with the instrument VLT/SPHERE \citep{Beuzit2008} as part of the SpHere INfrared survey for Exoplanets {\citep[SHINE,][]{Chauvin2017b}}. Our goals are to further characterize the orbital motion and parameters of HR~2562~B. A spectrophotometric analysis of the companion is presented in {\citet{Mesa2018}}. We describe the observations and the data reduction (Sect.~\ref{sec:data}). Subsequently, we use the new astrometry of HR~2562~B to confirm its companionship and analyze its orbital motion jointly with the GPI astrometry (Sect.~\ref{sec:astrometry}). We subsequently fit the SPHERE and GPI astrometry to derive first constraints on the companion's orbit (Sect.~\ref{sec:orbit}). {We analyze its potential dynamical interactions with the host-star debris disk (Sect.~\ref{sec:dynamics}). Finally, we discuss potential formation scenarios for the companion, the possibility to estimate its dynamical mass, and further insights into the system that will be provided by further astrometric monitoring of the brown dwarf companion and ALMA observations of the disk.} 

\section{Observations and data analysis}
\label{sec:data}

We observed HR~2562 on 2016 December 12, 2017 February 7, and 2017 September 29 with the SPHERE near-infrared (NIR) {camera IRDIS \citep{Dohlen2008a} and integral field spectrometer IFS} \citep{Claudi2008} simultaneously (Table~\ref{tab:obs}). For the first two epochs, the IRDIS data were acquired in the $H$-band broad-band filter ($\lambda_{H}$\,=\,1.6255~$\muup$m) with the aim to image the debris disk and IFS in $YJ$ mode (0.95--1.35~$\muup$m). As the disk was not detected with this setup, for the latest epoch, we used the IRDIFS\_EXT mode, that is, IRDIS with the $K12$ narrow-band filter pair \citep[$\lambda_{K1}$\,=\,2.110~$\muup$m and $\lambda_{K2}$\,=\,2.251~$\muup$m,][]{Vigan2010} and IFS covering the $YJH$ bands (0.95--1.65~$\muup$m). The star was imaged with an apodized pupil Lyot coronagraph {\citep{Carbillet2011, Martinez2009}} of inner working angle 95~mas (December 2016 and February 2017 data) or 120~mas ({September} 2017 data). The observing conditions were poor for the first observation, but the companion could still be detected and its astrometry extracted from the IRDIS data. Observing conditions were good to average for the second and third epochs. For calibrating the flux and the centering of the images, we acquired at the beginning and end of the sequences unsaturated non-coronagraphic images of the star (hereafter point-spread function or PSF) and coronagraphic images with four artificial crosswise replica of the star \citep{Langlois2013}. For the third sequence, the stellar replica were used for the whole sequence to minimize the centering errors in the astrometric error budget. Other calibration data (sky backgrounds, darks, detector flats) were obtained after the observations or during the daytime. 

\begin{table*}[t]
\caption{SPHERE astrometry relative to the star of HR~2562~B.}
\label{tab:astrometry}
\begin{center}
\begin{tabular}{l c c c c c c c}
\hline\hline
Epoch & Filter & $\rho$ (mas) & PA (deg) & $\Delta$RA (mas) & $\Delta$Dec (mas) & Pixel scale (mas/pix) & North correction angle ($^{\circ}$) \\
\hline
\multicolumn{8}{c}{IRDIS} \\
\hline
2016.95 & $H$ & {637.8$\pm$6.4} & {297.81$\pm$0.54} & {$-$564.1$\pm$4.9} & {297.6$\pm$4.1} & 12.251$\pm$0.009 & $-$1.808$\pm$0.043\\
2017.10 & $H$ & {644.0$\pm$2.3} & {297.82$\pm$0.19} & {$-$569.6$\pm$1.8} & {300.5$\pm$1.4} & 12.251$\pm$0.009 & $-$1.712$\pm$0.058\\
2017.75 & $K1$ & 661.2$\pm$1.3 & 297.97$\pm$0.16 & $-$583.9$\pm$1.1 & 310.1$\pm$0.8 & 12.267$\pm$0.009 & $-$1.735$\pm$0.043\\
2017.75 & $K2$ & 658.9$\pm$1.6 & 298.08$\pm$0.17 & $-$581.4$\pm$1.2 & 310.2$\pm$1.0 & 12.263$\pm$0.009 & $-$1.735$\pm$0.043\\
\hline
\multicolumn{8}{c}{IFS} \\
\hline
2017.10 & $YJ$ & 643.8$\pm$3.2 & 297.51$\pm$0.28 & $-$571.0$\pm$2.7 & 297.4$\pm$1.8 & 7.46$\pm$0.02 & $-$102.19$\pm$0.11\\
2017.75 & $YJH$ & 657.5$\pm$2.6 & 297.65$\pm$0.21 & $-$582.4$\pm$2.1 & 305.1$\pm$1.5 & 7.46$\pm$0.02 & $-$102.22$\pm$0.11\\
\hline
\end{tabular}
\end{center}
\tablefoot{The astrometric error bars were derived assuming an error budget including the measurement and systematic errors. The uncertainties in the estimation of the location of the star were derived to be 2.7 and 0.94~mas for the December 2016 and February 2017 IRDIS data sets, respectively. For the February 2017 IFS data set, this uncertainty is estimated to be 0.11~mas. The {September} 2017 data sets were acquired simultaneously with the satellite spots in the field of view.}
\end{table*}

\begin{figure}[t]
\centering
\includegraphics[trim = 0mm 0mm 0mm 0mm,clip,width=0.242\textwidth]{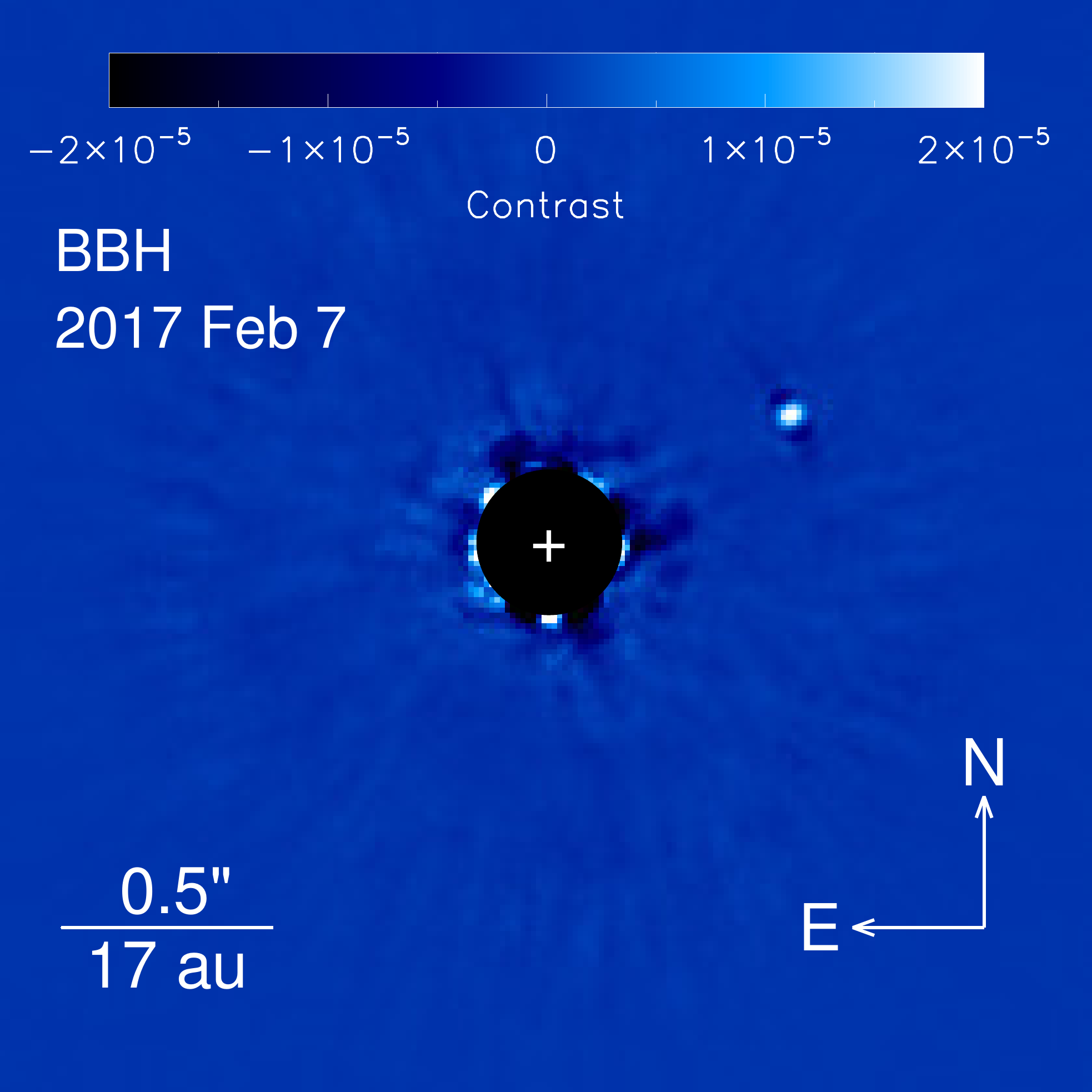}
\includegraphics[trim = 0mm 0mm 0mm 0mm,clip,width=0.242\textwidth]{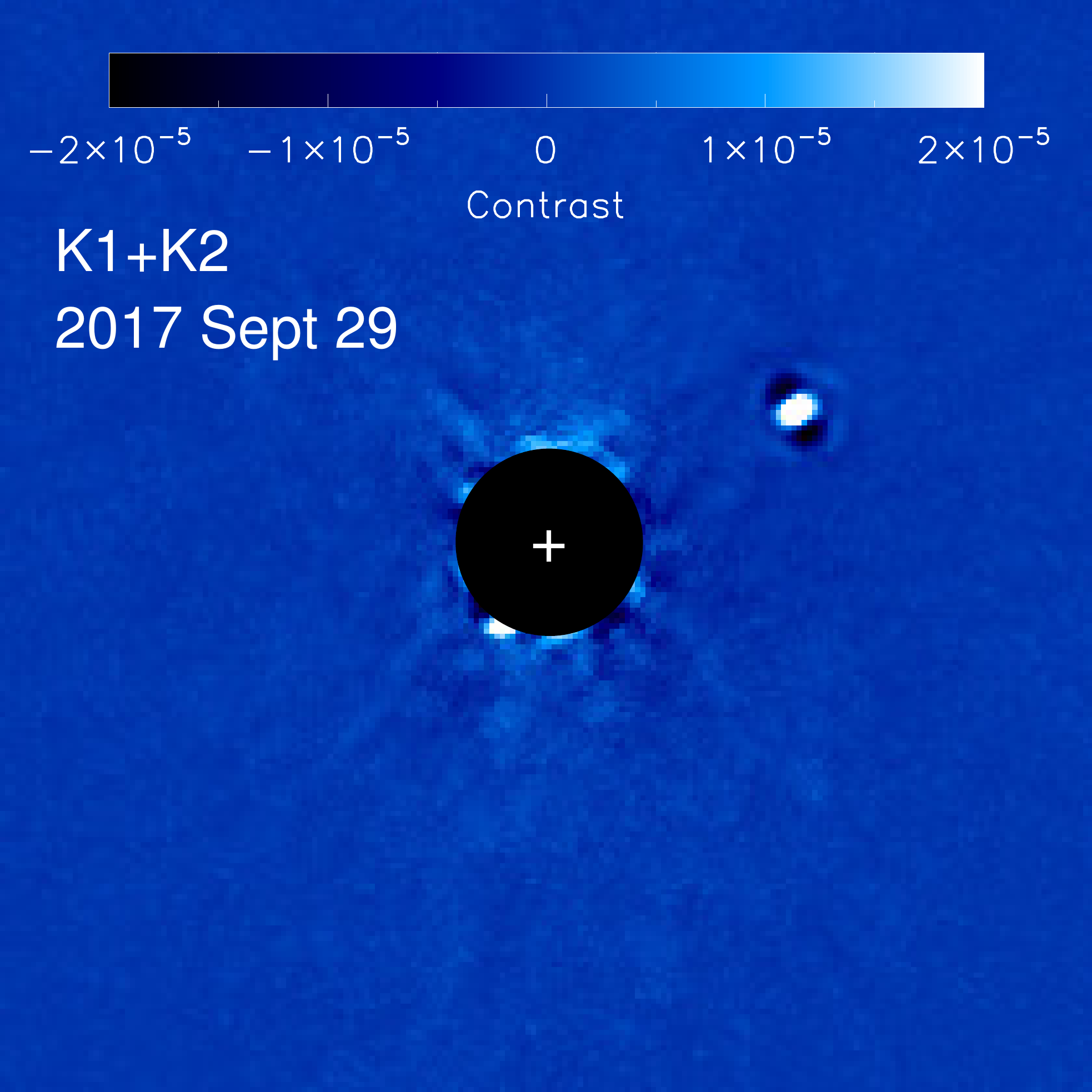}
\caption{{SPHERE/IRDIS TLOCI images of HR~2562 {obtained in the broad $H$-band filter ($\lambda_{H}$\,=\,1.6255~$\muup$m, \textit{left}) and with the combination of the narrow-band $K12$ filter pair images ($\lambda_{K1}$\,=\,2.110~$\muup$m, $\lambda_{K2}$\,=\,2.251~$\muup$m, right)}. The central regions of the images were numerically masked out to hide bright stellar residuals. The white crosses indicate the location of the star.}}
\label{fig:sphereim}
\end{figure}

The data were reduced with the SPHERE Data Center\footnote{\url{http://sphere.osug.fr/spip.php?rubrique16&lang=en}} pipeline {\citep{Delorme2017b}}, which uses the Data Reduction and Handling software \citep[v0.15.0,][]{Pavlov2008} and additional routines for the IFS data reduction \citep{Mesa2015}. The pipeline corrects for the cosmetics and instrument distortion, performs the wavelength calibration, and extracts the IFS image cubes, registers the frames, and normalizes their flux. Then, we sorted the frames using visual inspection and the statistics of their flux and selected about 60\% to 90\% of the best frames according to the data set. For the second IRDIS data set, we subsequently binned it temporally by a factor of two to avoid long computing times during the data post-processing while keeping the azimuthal smearing of the companion negligible. After these steps, for the IRDIS science cubes, we were left with 159, 129, and 92 frames, respectively. Finally, the data were analyzed with a consortium image processing pipeline \citep{Galicher2018}. We show in Fig.~\ref{fig:sphereim} median-collapsed contrast IRDIS images obtained with {Template Locally Optimized Combination of Images algorithm \citep[TLOCI,][]{Marois2014}}. For the IFS data, only the two last data sets could be used for extracting the companion astrometry. For the data analysis, we kept 63 and 50 frames, respectively. 

The known brown dwarf companion is detected at all epochs. Its astrometry was measured using TLOCI applied to each spectral channel of the science cubes separately. To attenuate the stellar residuals in an image, TLOCI subtracts from the image a model image of the stellar residuals built using the frames obtained in the same observing sequence. To account for the local properties of the stellar residuals, this model image or reference image is computed for each frame in a science cube in annuli with a width of 1.5 times the full width at half maximum (FWHM), and divided into sectors. To avoid large photometric and astrometric biases on putative point sources, the reference images were built using the best linear combination of the 80 most correlated frames for which the self-subtraction of mock point sources, modeled using the observed PSF, was at maximum 15\% (December 2016 data set) and 30\% (February and {September} 2017 data sets). To accurately estimate the astrometry and photometry of HR~2562~B while accounting for the TLOCI biases, we created a science cube with only a mock companion modeled from the observed PSF inserted at the rough location (within a pixel accuracy) of the measured companion accounting for the field-of-view rotation \citep{Galicher2011a}. We then processed the data with TLOCI assuming the algorithm coefficients computed for the analysis without the mock companion. After, the subpixel position and flux of the model companion image were optimized to minimize the image residuals within a disk of radius 1.5~FWHM centered on the measured companion. The astrometry reported in Table~\ref{tab:astrometry} was calibrated following the methods described in \citet{Maire2016b}. We compared the IRDIS positions of stars in fields in 47~Tuc and NGC~3603 to HST positions \citep[A. Bellini, priv. comm.;][]{Khorrami2016} to determine the pixel scale and the correction angle to align the images with the North direction. Since the IFS observations are performed simultaneously with the IRDIS observations, we calibrated the IFS data of HR~2562 using the IRDIS calibration and an additional angle offset accounting for the relative orientation between the two instrument fields of view. We compared the TLOCI astrometry to the results from the {ANgular DiffeRential OptiMal Exoplanet Detection Algorithm \citep[ANDROMEDA,][]{Mugnier2009, Cantalloube2015}} and a {principal component analysis algorithm} \citep{Mesa2015}. All values are compatible given the error bars and we decided to use the IRDIS astrometry extracted with TLOCI in the $H$ and $K1$ bands for the astrometric and orbital analyses.

\section{Astrometric confirmation and orbital motion}
\label{sec:astrometry} 

\begin{figure}[t]
\centering
\includegraphics[trim = 8mm 4mm 2mm 10mm, clip, width=0.4\textwidth]{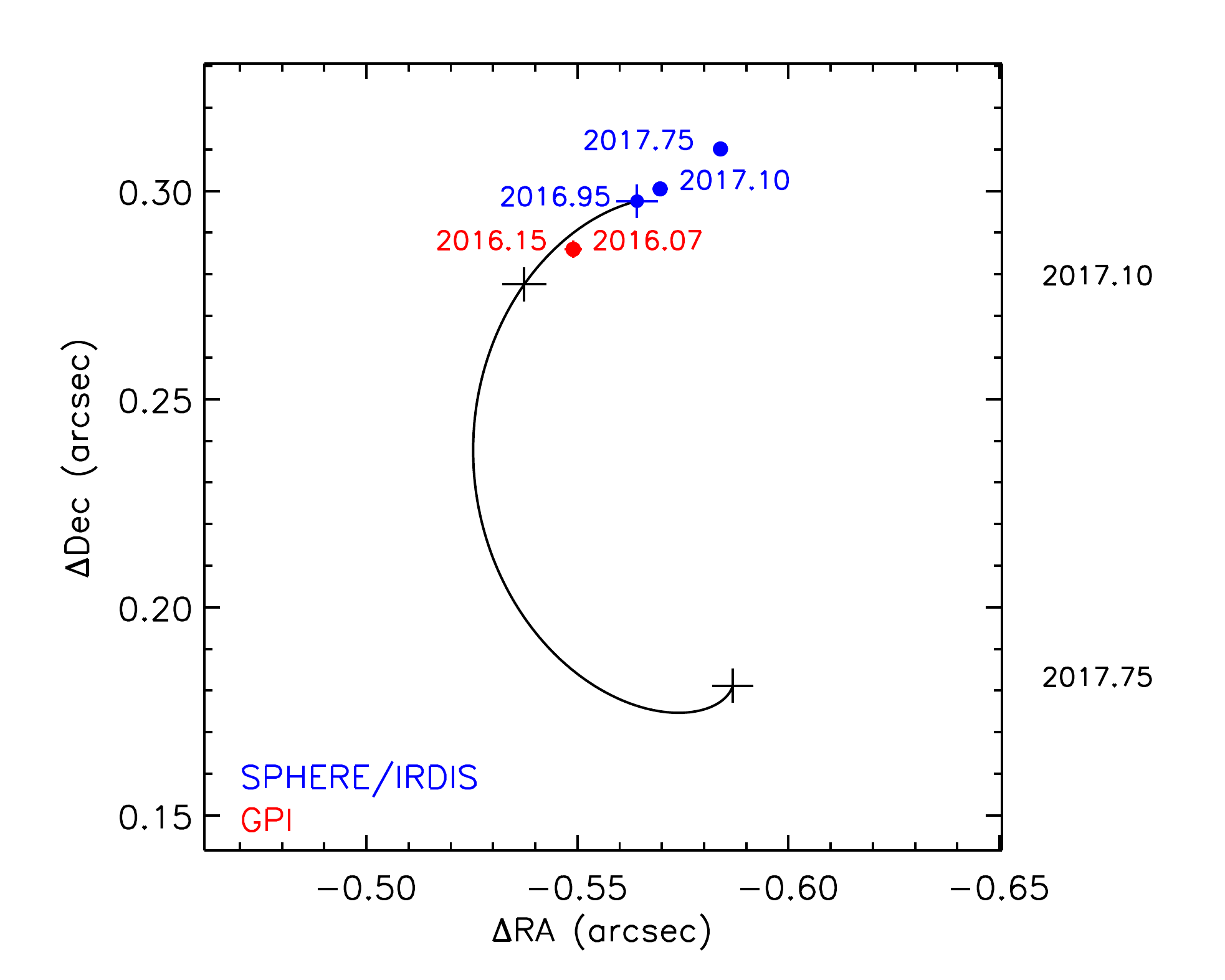}
\caption{{SPHERE relative astrometry of HR~2562~B (blue points). The black curve shows its motion if it is a stationary background object. The black crosses represent the locations at epochs 2017.10 and 2017.75 (see labels on the right side outside the plot) under the stationary background hypothesis accounting for the uncertainties in the stellar proper motion and distance. The GPI astrometry (red points) is shown for comparison. {For most of the data points, the uncertainties} are smaller than the size of the symbols.}}
\label{fig:cpmtest}
\end{figure}

We show in Fig.~\ref{fig:cpmtest} the common proper motion test of the companion. Already considering only the December 2016 and February 2017 epochs, the companion does not follow the stationary background track at {5.7~$\sigma$ in right ascension and 5.2~$\sigma$ in declination}. The addition of the September 2017 epoch reveals a significant orbital motion for the companion (see below) that is not consistent with the motion expected for a stationary background object.

\begin{figure*}[t]
\centering
\includegraphics[trim = 1mm 10mm 0mm 5mm,clip,width=0.96\textwidth]{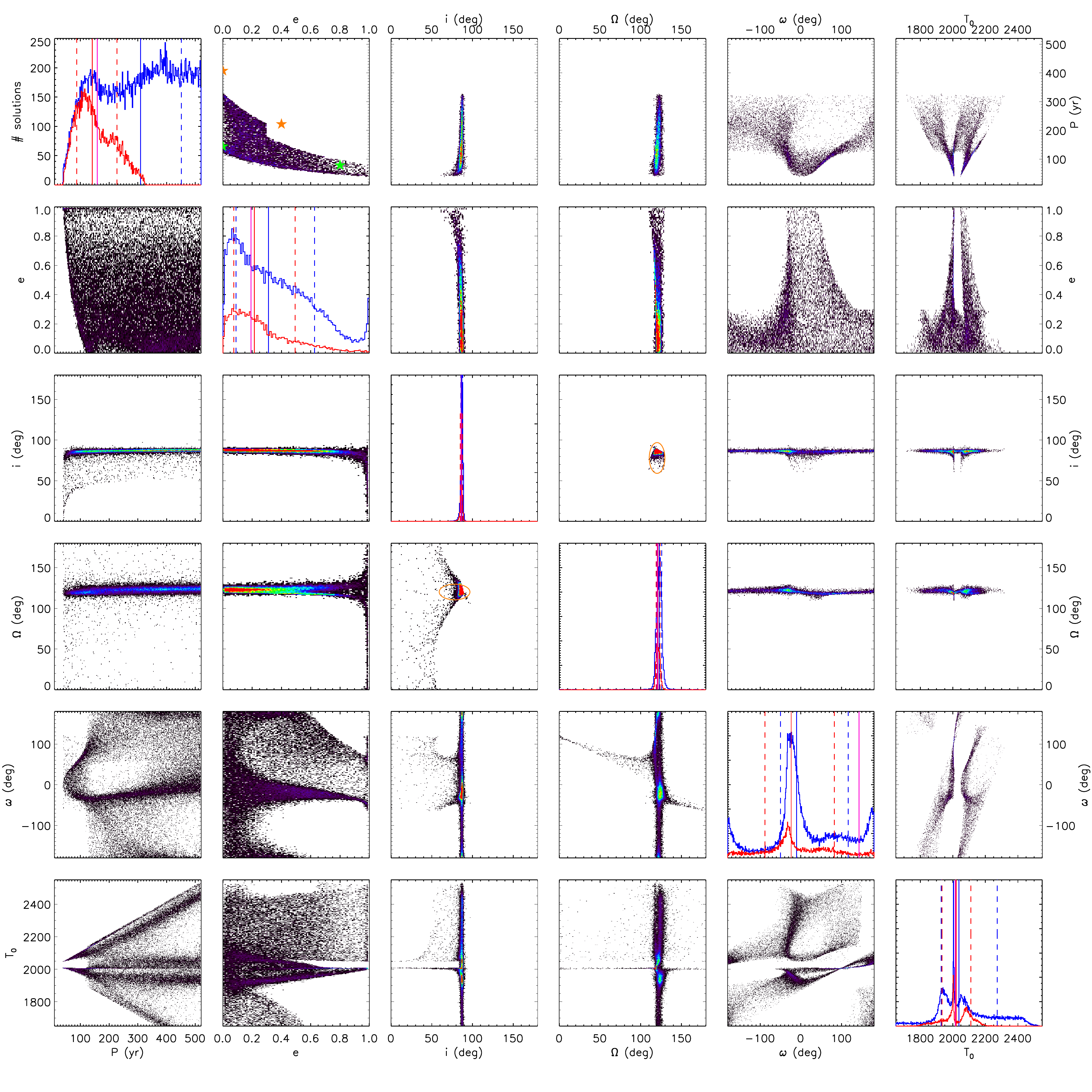}
\caption{{LSMC distributions of the six {Campbell} orbital elements for all the fitted solutions with $\chi_{\rm{red}}^2$\,<\,2 among 2\,000\,000 random trials. The diagrams displayed on the diagonal from top left to lower right represent the 1D histograms for the individual elements (blue: all solutions, red: {solutions compatible with the estimated disk geometry at 3~$\sigma$ and disk cavity size}). The off-diagonal diagrams show the correlations between pairs of orbital elements, with diagrams below and to the left of the diagonal showing all the fitted solutions, and the diagrams above and to the right of the diagonal only showing solutions which are {compatible with the estimated disk geometry and cavity size} (see Sect.~\ref{sec:dynamics}). The linear color scale in the correlation plots account for the relative local density of orbital solutions. In the histograms, the {purple and magenta solid lines indicate the best $\chi^2$ fitted solutions for all solutions and the disk-compatible solutions, respectively. The solid and dashed lines of a given color show the 50\% percentile values and the intervals at 68\% (blue: all solutions, red: disk-compatible solutions). The orange ellipses in the $i$-$\Omega$ plots show the disk inclination and position angle estimated by \citet{Moor2015} at 3~$\sigma$.} The stars in the eccentricity-period diagram for the restricted solutions (top row, second panel from the left) indicate the configurations tested in the N-body simulations described in Sect.~\ref{sec:dynamics} (green: allowed, orange: excluded).}}
\label{fig:cornerplot}
\end{figure*}

Subsequently, we combined the SPHERE/IRDIS astrometry with the GPI data reported in \citet{Konopacky2016b} to analyze the companion's orbital motion. The total time baseline of the measurements represents {$\sim$1.7~yr}. With respect to the last GPI epoch (late February 2016), the separation of the companion in early February 2017 increased by $\sim$25~mas at $\sim$7--8~$\sigma$ significance with a current orbital motion projected in the plane of the sky of $\sim$25~mas ($\sim$0.85~au) per year, whereas its position angle does not show any significant variations (see Sect.~\ref{sec:orbit}). The strong increase in separation is too large to be accounted for by small systematic errors between the SPHERE and GPI astrometry. The separation measured in the SPHERE {September} 2017 data confirms the observed trend (increase of $\sim$15~mas with respect to February 2017). The large separation increase also rules out a face-on circular orbit. For the position angles, we could not exclude {small systematic errors between SPHERE and GPI} when considering the 2016 and early 2017 data points. Nevertheless, the SPHERE {September} 2017 data point confirms that the observed evolution for the position angle is genuine.
\begin{figure*}[t]
\centering
\includegraphics[trim = 8mm 4mm 2mm 4mm,clip,width=0.4\textwidth]{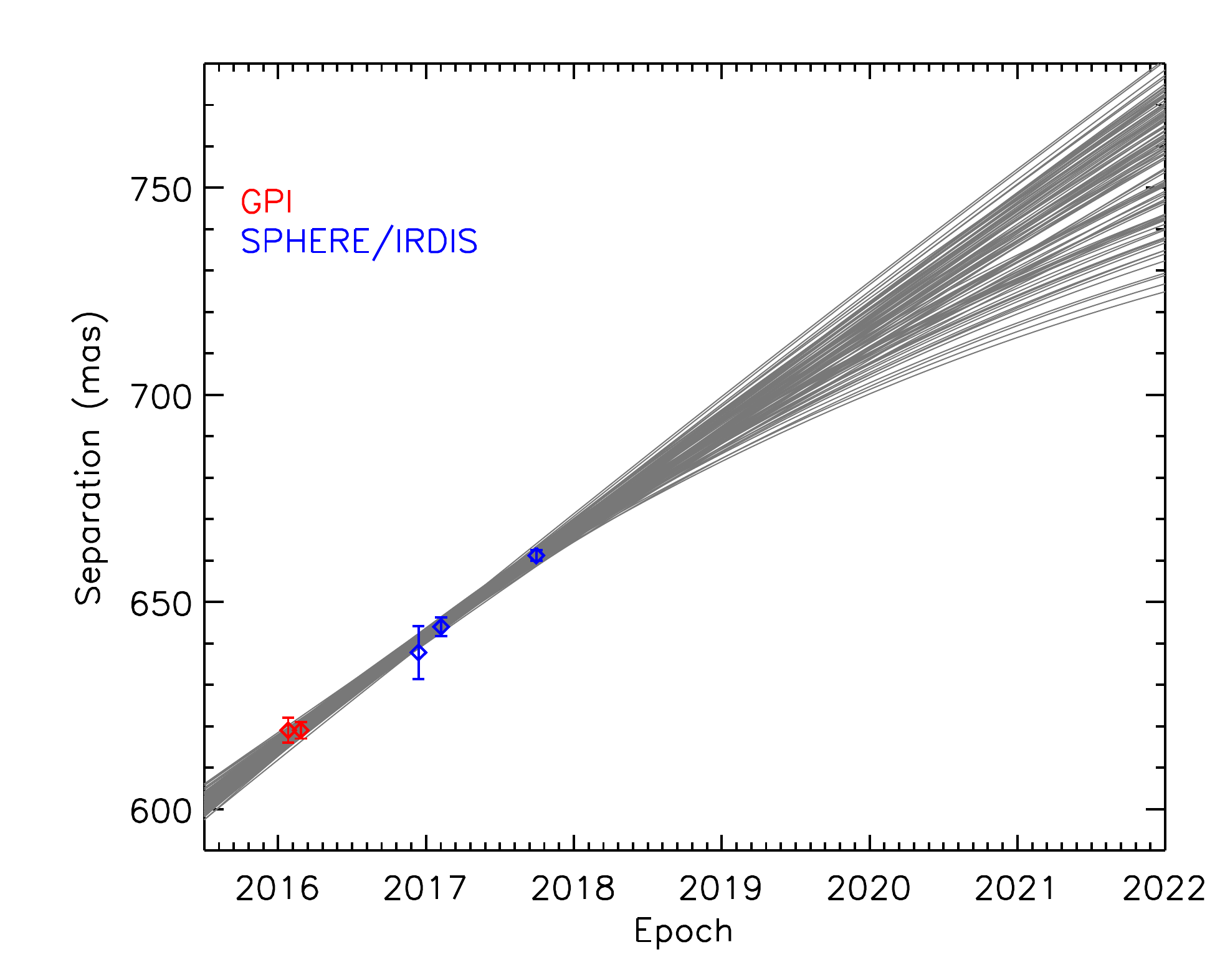}
\includegraphics[trim = 8mm 4mm 2mm 4mm,clip,width=0.4\textwidth]{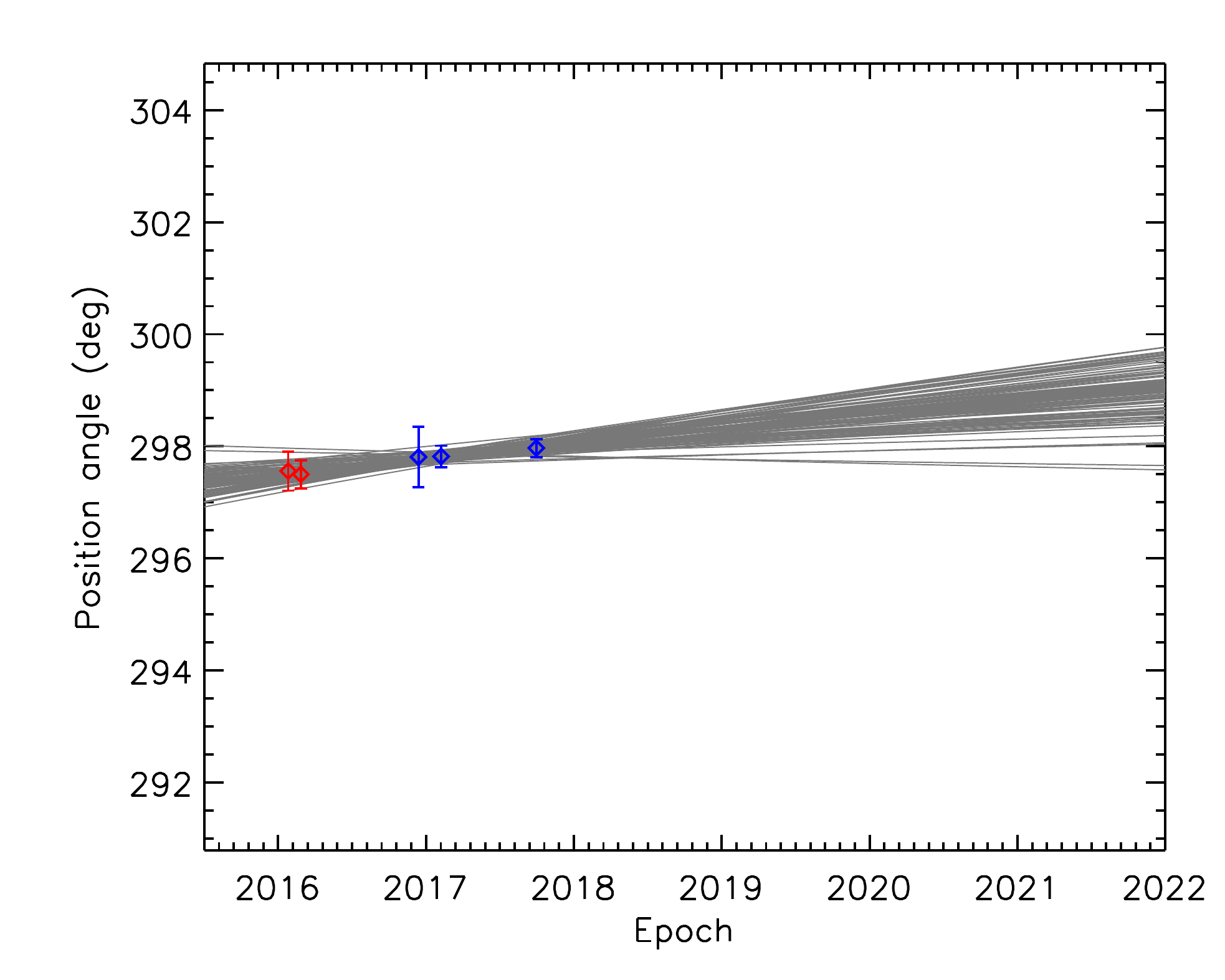}
\caption{{Temporal evolution of the separation and position angle of HR~2562~B measured by GPI and SPHERE. Predicted separations and position angles for 100 randomly selected orbital solutions in the upper-right part of Fig.~\ref{fig:cornerplot} are also shown.}}
\label{fig:seppatimepredic}
\end{figure*}

\section{Orbital fitting}
\label{sec:orbit}

We used a least-square Monte Carlo (LSMC) procedure to fit the SPHERE and GPI astrometry \citep{Esposito2013,Maire2015}. We assumed for the system the \textit{Gaia} distance and a total mass of 1.3~$M_{\sun}$. We drew 2\,000\,000 random realizations of the astrometric measurements assuming Gaussian distributions around the nominal values, and then fitted the six {Campbell} elements simultaneously using a debugged version of the downhill simplex \texttt{AMOEBA} algorithm\footnote{The own built-in routine provided by {the Interactive Data Language (IDL) programming language} truncates the stepping scales to floating point precision, regardless of the input data type.} \citep{Eastman2013}: orbital period $P$, inclination $i$, longitude of node $\Omega$, argument of periastron passage $\omega$, and time at periastron passage $T_0$. Initial guesses for the orbital elements were drawn assuming uniform distributions. Given the limited orbital coverage of the data, we considered two cases: (1) no priors on the orbital elements except for the period ($P$\,=\,10--2000\,yr), and (2) orbits {with the same period prior and} coplanar with the debris disk measured with \textit{Herschel} \citep[$i$\,$\sim$\,78.0$\pm$6.3$^{\circ}$, $\Omega$\,$\sim$\,120.1$\pm$3.2$^{\circ}$,][]{Moor2015}. We found that without any prior, the orbital solutions clearly favor a coplanar configuration with the disk. To test the presence of biases in the fitted eccentricity and time at periastron passage because of the small covered orbital arc, we used the correction proposed by \citet{Konopacky2016a} but did not find large differences between the derived distributions and we decided to keep the non-corrected distributions for the analysis.

The lower-left part of Fig.~\ref{fig:cornerplot} shows the histogram distributions and the correlation diagrams of the orbital parameters for the case without using the disk measured inclination and position angle constraints for all the derived solutions with $\chi_{\rm{red}}^2$\,<\,2. The 68\% intervals for the parameters are: {$e$\,$\sim$\,0.09--0.63, $i$\,$\sim$\,86--88$^{\circ}$, and $\Omega$\,$\sim$\,119--126$^{\circ}$. The period is unconstrained. The distribution for the argument of periastron exhibits two peaks around $-$25$^{\circ}$ and at $\sim$175$^{\circ}$. The distribution for the time at periastron passage shows a very narrow peak in $\sim$2000 with two broader side peaks in $\sim$1940 and $\sim$2050. 

We compared these results with those from a Markov-chain Monte Carlo (MCMC) tool \citep{Chauvin2012} assuming uniform priors in log\,$P$, $e$, cos\,$i$, $\Omega + \omega$, $\omega - \Omega$, and $T_0$. We found similar ranges for the inclination, the longitude of node, and the argument of periastron. However, the MCMC period and time at periastron distributions are better defined and the corresponding eccentricity distribution shows a very strong peak close to $e$\,=\,1. A high-eccentricity peak is also seen in the LSMC distribution but significantly weaker. Additional checks showed that the better constraints on the period and time at periastron as well as the strong high-eccentricity peak obtained in the MCMC analysis are related to the period prior. {The high-eccentricity peak feature can be explained by the almost radial motion without curvature and the absence of significant change in position angle of the companion over the time baseline. We also note that this almost radial motion of the companion results in the well-constrained orbit plane derived in the orbital fits.}

We also used the imorbel online tool\footnote{\url{http://drgmk.com/imorbel/}} to apply the small arc analysis of \citet{Pearce2015}. We derived {68\% intervals of 0.16$^{+0.02}_{-0.02}$ for the dimensionless parameter $B$ and {6.99$^{+3.33}_{-3.42}$~deg for the angle $\phi$ between the projected separation and velocity vectors of the companion}. Using these values and Figs. 5 and 6 in \citet{Pearce2015}, we can set constraints on the minimum semi-major axis, minimum eccentricity, and maximum inclination for the companion. The minimum semi-major axis is 13$^{+1}_{-2}$~au at 3~$\sigma$. (Quasi-)circular and/or edge-on orbits cannot be excluded.} These constraints are compatible with the LSMC results, but the latter are more stringent for the inclination.

\citet{Moor2015} derived from a geometrical model fitted to \textit{Herschel}/PACS data an average dust radius of 112.1$\pm$8.4~au, an inner hole radius of $\sim$38$\pm$20~au, and an average outer dust radius of $\sim$187$\pm$20~au. Because of inconsistencies between the fitted values of the average dust radius between the \textit{Herschel}/PACS images and the SED \citep[64$\pm$6~au,][]{Moor2015}, \citet{Konopacky2016b} performed a simultaneous fit of the SED and \textit{Herschel} image and derived an inner hole radius of $\sim$75~au. From this constraint and assuming a circular orbit for the companion, \citet{Konopacky2016b} derived an upper mass limit for the companion of $\sim$0.24~$M_{\odot}$, which is much larger than the upper limit from evolutionary models (when using an inner hole radius of 38~au, the value is $\sim$20~$M_{\rm{J}}$). \citet{Konopacky2016b} propose that this apparent discrepancy for the companion mass could be solved if the companion has an eccentric orbit. This hypothesis is compatible with the results of our orbital analysis, although new observations are required to obtain robust constraints.

Figure~\ref{fig:seppatimepredic} represents the predicted separations and position angles for 100 randomly selected fitted orbits {compatible with the disk geometry and cavity size constraints} (see Sect.~\ref{sec:dynamics}). Monitoring the system in subsequent years will be critical for improving the orbital constraints, especially if orbital curvature can be measured. {We note that, because of the small baseline of the measurements, the separations of the most extreme orbital predictions diverge quickly with time after the last SPHERE epoch and that the maximum difference in separations is already {$\sim$15~mas} in early 2019. A significant deviation from linearity could therefore be measured, that would favor short-period orbits with non-zero eccentricities over long-period and circular orbits. If this is the case, the {robust rejection of circular orbits will however require a longer follow-up}. If no or small deviation from linearity is measured, this would reject a few short-period orbits and bring only little improvement on the derived orbital elements. We also emphasize that our orbital constraints are preliminary and that robust constraints will be possible once at least one-third of the complete companion orbit can be monitored.}

Finally, we used the methods in \citet{Pearce2014} to test the scenario of an unseen inner low-mass companion which could bias the orbital eccentricity of HR~2562~B toward large values due to the orbital motion that the unseen companion induces on the host star. For this, we used the period and eccentricity distributions derived from the orbital fit for the non-restricted case. Figure~\ref{fig:pearcetest} shows the mass of a putative inner companion as a function of the eccentricity of HR~2562~B. Such a companion would lie at an angular separation of 0.1$''$. We estimated the TLOCI contrast limit in the SPHERE/IRDIS February 2017 data set at this separation including the coronagraph transmission \citep{Boccaletti2018} and the small sample statistics correction \citep{Mawet2014} and derived a value of $\sim$1.6$\times$10$^{-3}$. This corresponds to a mass of $\sim$0.1~$M_{\odot}$ for a system age of 450~Myr according to the evolutionary and atmospheric models of \citet{Baraffe2003, Baraffe2015}. Thus, we can conclude that if HR~2562~B has an eccentricity larger than $\sim$0.6, this eccentricity is genuine and does not result from an unseen low-mass inner companion. For smaller eccentricities, we cannot exclude an unseen inner companion as potential origin. As discussed in {\citet{Mesa2018}}, there is no clear evidence for binarity of the host star, although additional observations are required to definitely rule out this hypothesis.

\section{Disk-companion dynamical analysis}
\label{sec:dynamics}

\subsection{Empirical and numerical dynamical analysis}

The positions we observed for HR~2562~B represent a very small part of its orbit, and the orbital fit is thus not able to give strong constraints. \citet{Konopacky2016b} showed that the companion mass is consistent with a stirring of the dust up to the outer edge of the disk given the constraints on the system age. By simulating dynamical interactions between the companion and the debris disk, we can further constrain the companion's orbit by removing solutions that do not match the observational constraints on the disk (e.g., measured cavity size, resolved image).

A word of caution about the accuracy on the disk parameters estimated by \citet{Moor2015} is needed here as the disk is only marginally resolved in the \textit{Herschel} data because of a large instrument PSF. The \textit{Herschel} constraints must therefore be treated with caution, especially for the size of the cavity. The fact that the disk is not detected in the SPHERE images, despite being highly inclined, indicates a low surface brightness in the NIR, which in turn suggests a spatially extended disk.

\begin{figure}[t]
\centering
\includegraphics[trim=14mm 0mm 14mm 10mmm,clip,width=0.42\textwidth]{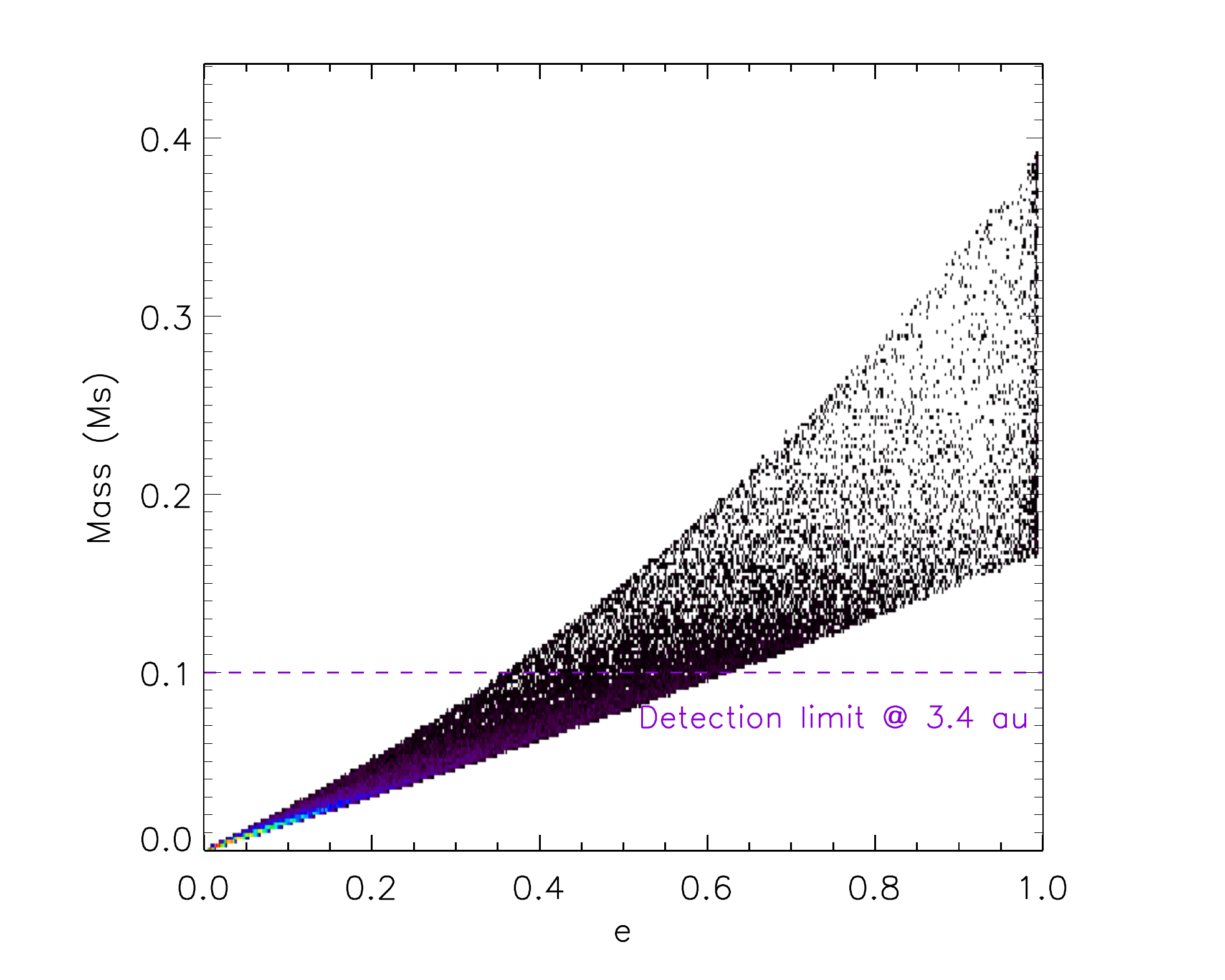}
\caption{{Mass (in solar masses) of an unseen inner companion that could bias the orbital eccentricity measured for HR~2562~B for the unrestricted case (lower-left part of Fig.~\ref{fig:cornerplot}) compared to the SPHERE/IRDIS detection limit at the separation predicted for this putative companion (purple line, see text).}}
\label{fig:pearcetest}
\end{figure}

{The distribution of the relative inclination of the companion orbit to the debris disk indicates that they are coplanar at the $\sim$1.5-$\sigma$ level (Fig.~\ref{fig:irelative}). Therefore, we} restricted our problem to orbits coplanar with the disk. We checked that a small misalignment of 20$^{\circ}$ between the companion orbit and the disk plane has little influence on the carving of the disk by the companion (Appendix~\ref{sec:dynamics_noncoplanar}).

\begin{figure}[t]
\centering
\includegraphics[width=.42\textwidth,trim = 4mm 4mm 13mm 8mm,clip]{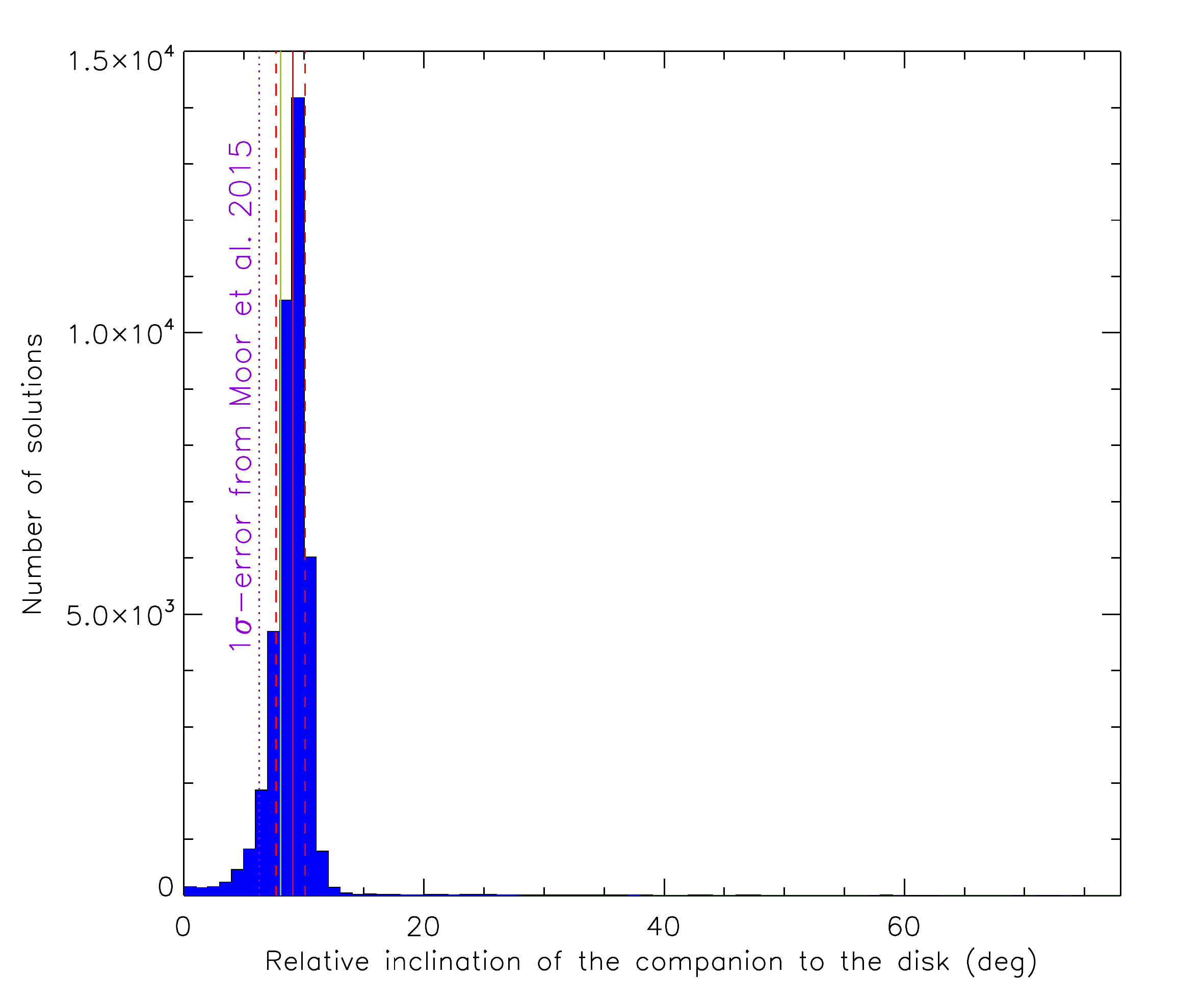}
\caption{{Relative inclination of HR~2562~B with respect to the debris disk from the unrestricted orbital fit given the constraints in \citet{Moor2015}. The red solid and dashed lines show the 50\% percentile value and the interval at 68\%, the green solid line shows the best-fit solution, and the purple dotted line indicates the 1$\sigma$ uncertainty on the inclination estimate in \citet{Moor2015}.}}
\label{fig:irelative}
\end{figure}

N-body simulations being time consuming, we first used empirical laws to obtain rough estimates of the parameter space of the orbital solutions compatible with the estimated disk cavity radius. The only relevant orbital parameters are the semi-major axis $a$ (or, equivalently, the period $P$) and the eccentricity $e$. Given these two parameters, we used the formulae in \citet{Wisdom1980} and \citet{Mustill2012} to compute the size of the gap opened by a companion following the approach of \citet{Lazzoni2018}. Unfortunately, the constraints on the disk gap radius are rather blurry, from 38$\pm$20~au to about 75~au. If we suppose that the disk gap radius is 75~au, we can still exclude \textit{a posteriori} a large part of the orbital fit results, as represented in the eccentricity-period panels in Fig.~\ref{fig:cornerplot}.

To test the validity of the empirical results, we then used the symplectic N-body code SWIFT\_RMVS3 \citep{Levison1994} to simulate the disk dynamics. The code does not simulate collisions between particles. {For the initial parameters of the particles, we assumed 10\,000 particles with a uniform distribution for their distance to the star between 1 and 200~au, hence their surface density is inversely proportional to their distance. Their eccentricity and their relative inclination to the disk $i_{\rm{r}}$ were drawn assuming uniform priors in eccentricity between 0 and 0.05 and in sin($i_{\rm{r}}$) between 0 and 2$^{\circ}$, respectively. We finally set the mass of the companion to 30~$M_{\rm{J}}$.} We checked with additional simulations that the mass assumed for the companion has little effect on the disk properties within the constraints from evolutionary models. The revolution of a companion within a debris disk is expected to first cause a gap in the dust distribution, that will be completely formed after 10\,000 companion revolutions \citep[see e.g.,][]{Holman1999}, and then the propagation of a spiral structure towards the outer edge of the disk, that will become more and more tightly wound because of the disk's secular precession and will eventually disappear \citep{Wyatt2005}. The age of the system is not well constrained either; \citet{Konopacky2016b} found estimates from 180~Myr to 1.6~Gyr in the literature \citep{Asiain1999, Rhee2007, Casagrande2011, Moor2011, Pace2013, Moor2015}, but concluded on an age range of 300--900~Myr. {\citet{Mesa2018}} determined a younger age upper limit of 750~Myr. Simulating a debris disk interacting with a companion for several hundred million years requires significant computing time; we therefore calculated the typical timescale $\tau$ of the dissipation of the wave from \citet{Wyatt2005}. We then set  the duration of our simulations to 100~Myr accordingly. This duration is supposed to correspond to a steady state. In fact, simulations revealed that the disk undergoes practically no change from an age of 10~Myr.

\begin{figure}[t]
\includegraphics[width=.24\textwidth,trim={0cm 1cm .5cm 2cm}]{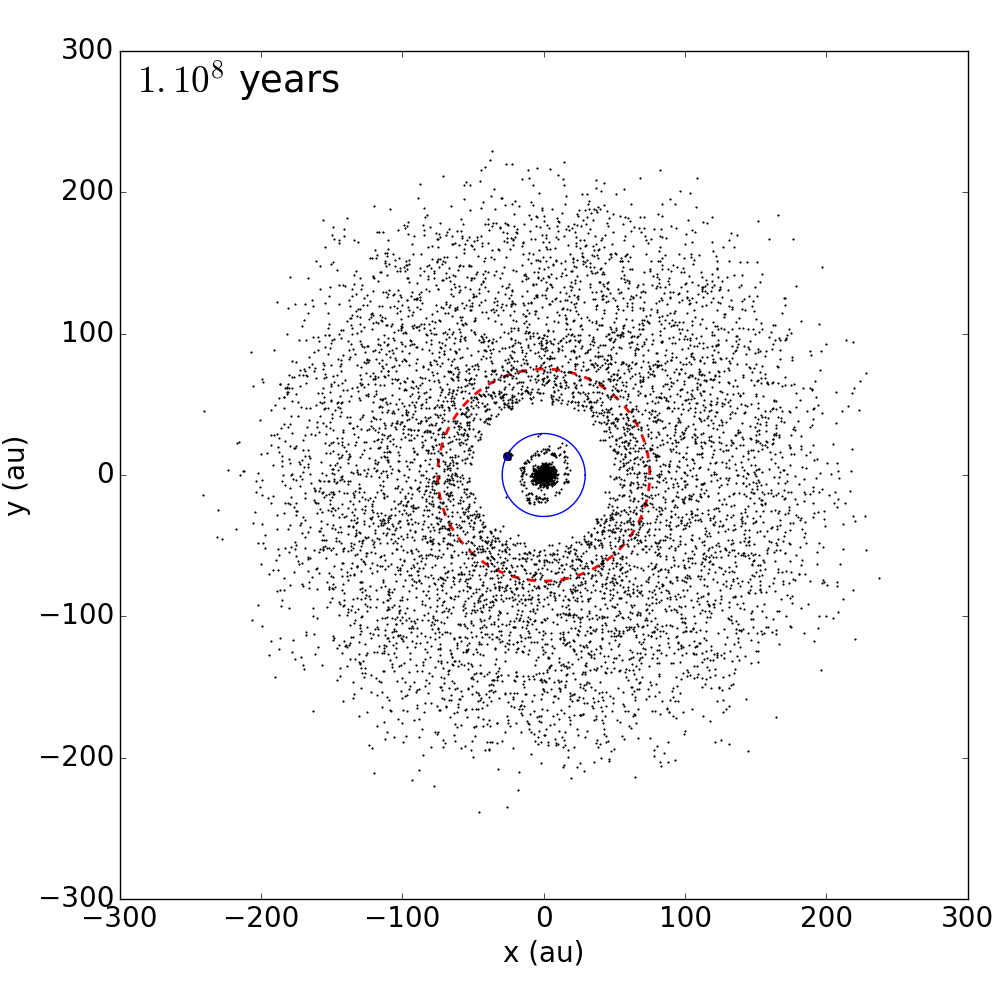}
\includegraphics[width=.24\textwidth,trim={0cm 1cm .5cm 2cm}]{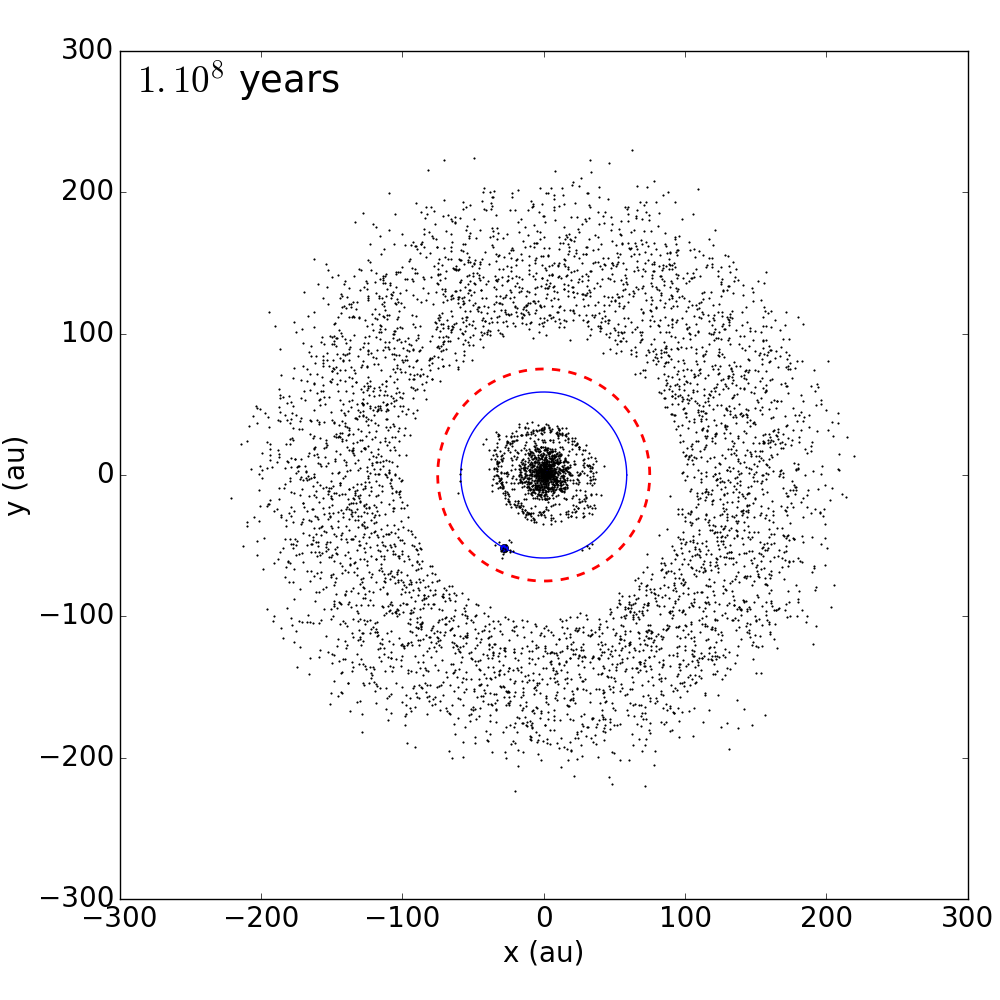}
\caption{N-body simulated images of the HR~2562 system after 100~Myr of evolution assuming a coplanar circular orbit for the companion with semi-major axes 30~au (\textit{left}) and 60~au (\textit{right}). The blue solid line shows the companion orbit and the red dashed line the maximum gap radius allowed by the observations.}
\label{fig:e=0}
\end{figure}

\begin{figure}[t]
\centering
\includegraphics[width=.24\textwidth,trim={0cm 1cm .5cm 2cm}]{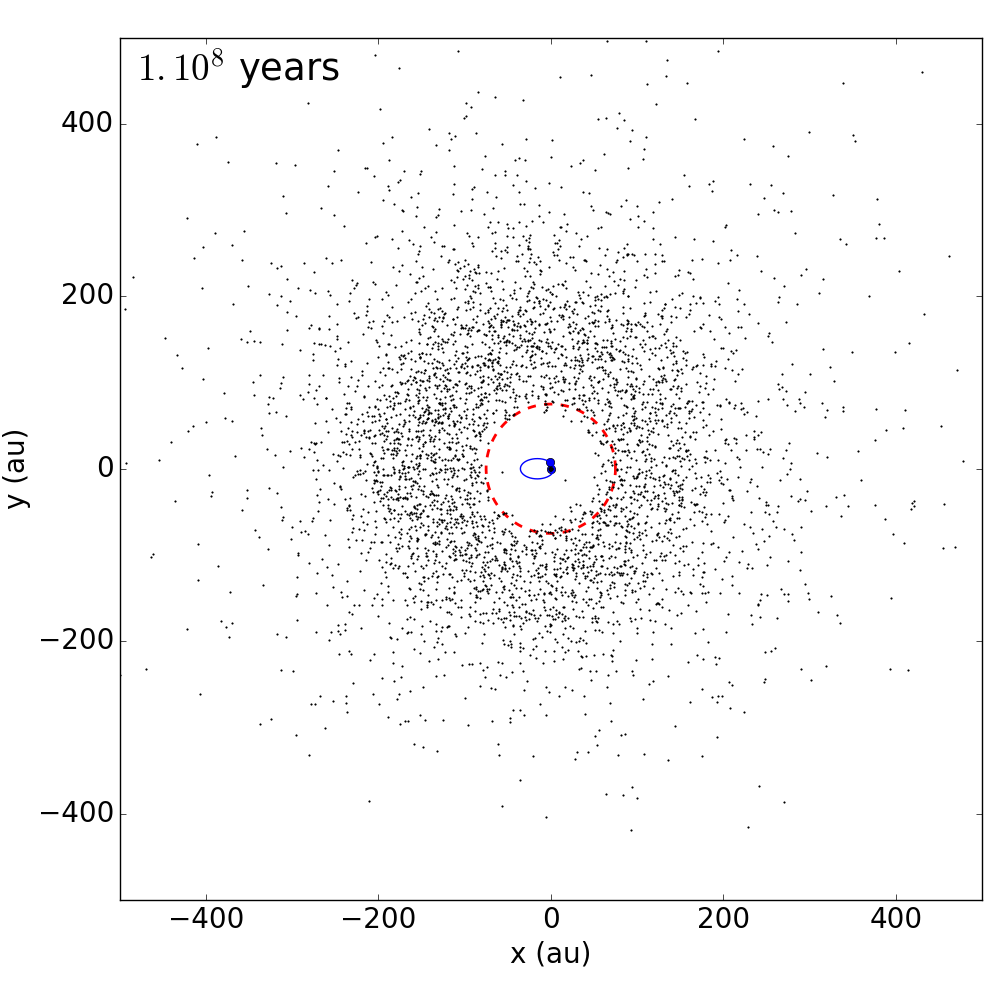}
\includegraphics[width=.24\textwidth,trim={0cm 1cm .5cm 2cm}]{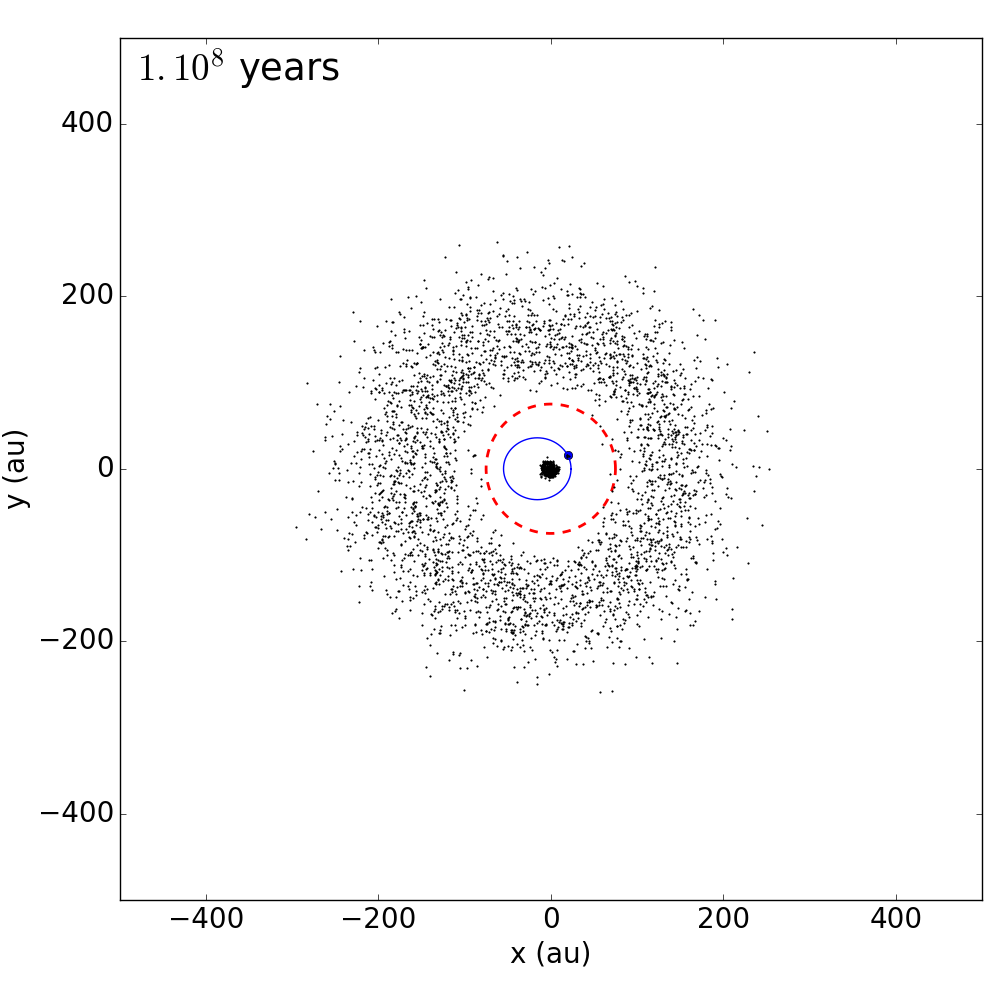}
\caption{As in Fig.~\ref{fig:e=0} but for eccentric orbits of the companion. \textit{Left}: $a$\,=\,20~au, $e$\,=\,0.8. \textit{Right}: $a$\,=\,40~au, $e$\,=\,0.4. {The diagrams have different} horizontal and vertical image cuts with respect to Fig.~\ref{fig:e=0}.}
\label{fig:e}
\end{figure}

\begin{figure*}[t]
\centering
{\begin{minipage}[c]{0.25\linewidth}
\centering
\subfloat[]{\includegraphics[width=0.75\linewidth]{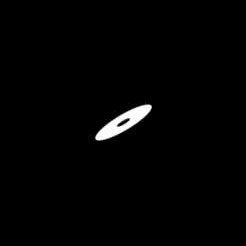}}
\end{minipage}
\begin{minipage}[c]{0.25\linewidth}
\centering
\subfloat[]{\includegraphics[width=.5\linewidth,trim={5.5cm 0 5.5cm 0}]{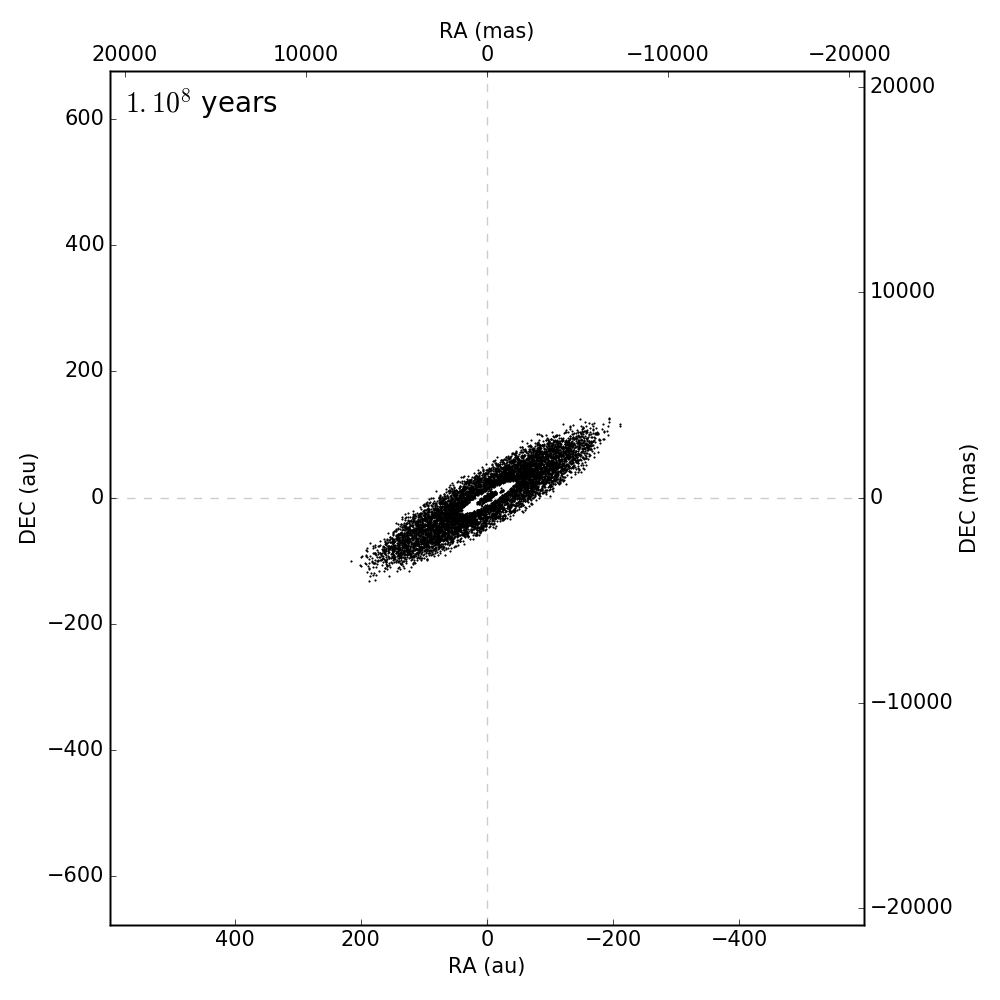}}
\end{minipage}
\begin{minipage}[c]{0.25\linewidth}
\centering
\subfloat[]{\includegraphics[width=.5\linewidth,trim={5.5cm 0 5.5cm 0}]{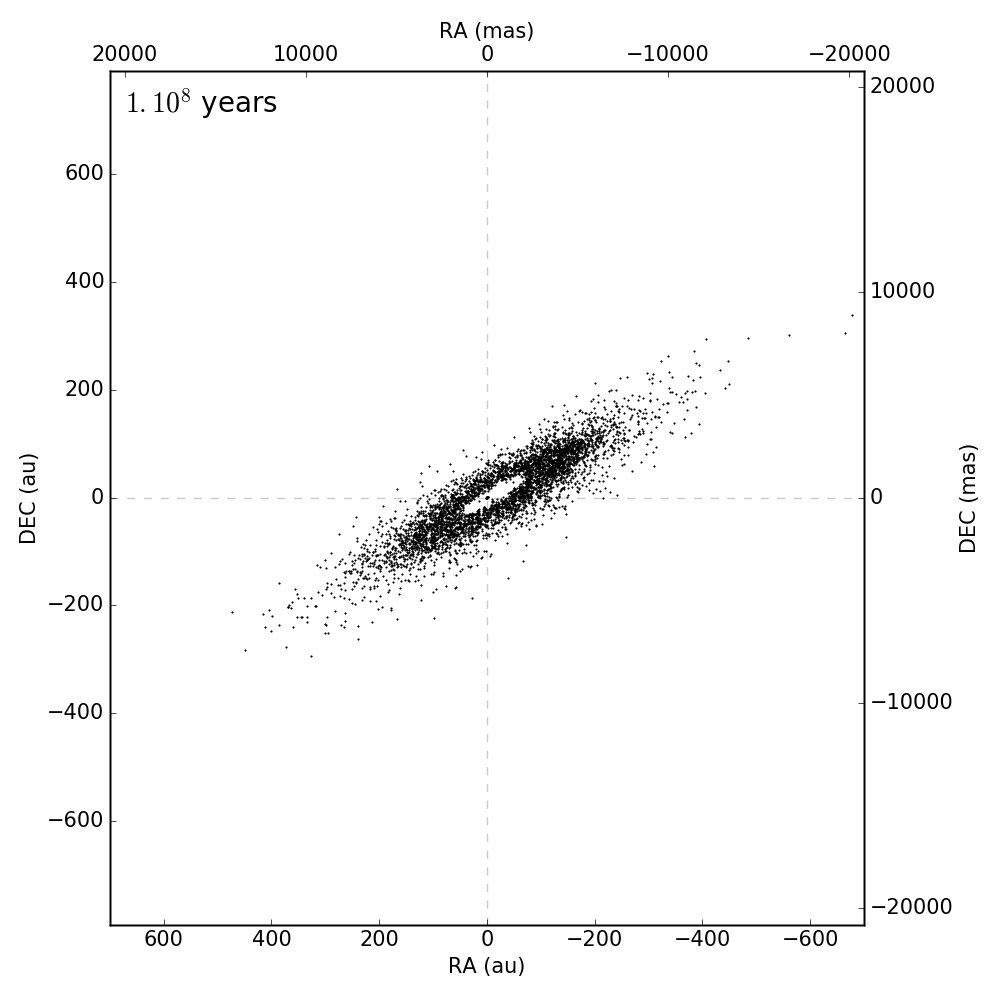}}
\end{minipage}
\vspace{.2cm}\\
\begin{minipage}[c]{0.25\linewidth}
\centering
\subfloat[]{\includegraphics[height=.8\linewidth,trim={0cm 0 0cm 0}]{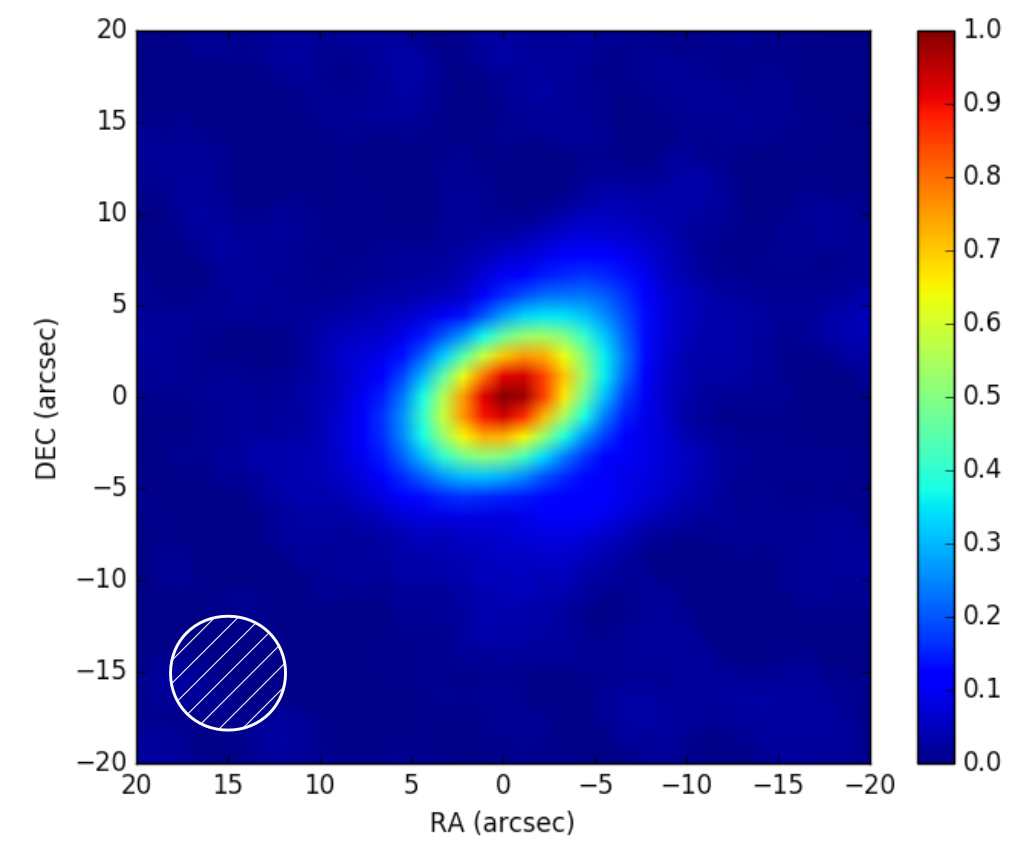}}
\end{minipage}
\begin{minipage}[c]{0.25\linewidth}
\centering
\subfloat[]{\includegraphics[height=.8\linewidth,trim={1cm 0cm 1cm 1cm}]{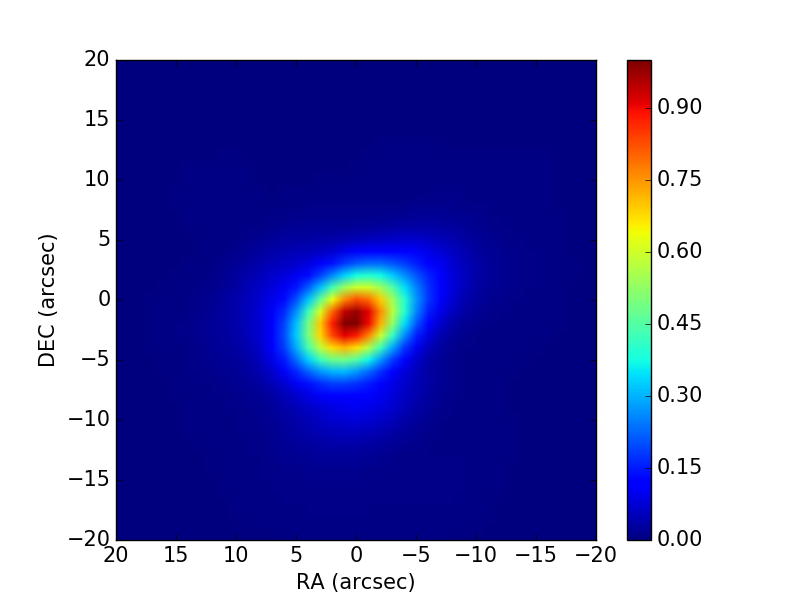}}
\end{minipage}
\begin{minipage}[c]{0.25\linewidth}
\centering
\subfloat[]{\includegraphics[height=.8\linewidth,trim={1cm 0cm 1cm 1cm}]{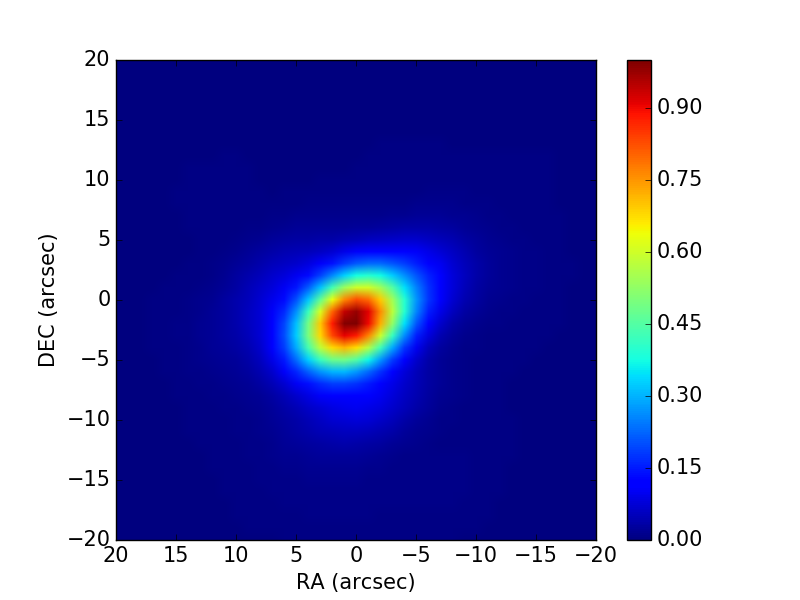}}
\end{minipage}}
\caption{(a) Geometrical fit to the \textit{Herschel}/PACS image at 70~$\mu$m adapted from \citet{Moor2015} shown in panel (d). (b) and (c) Simulated images consistent with the {estimated disk cavity for a circular configuration ($a$\,=\,30~au, $e$\,=\,0) and for an eccentric configuration ($a$\,=\,40~au, $e$\,=\,0.4), respectively}. (d) \textit{Herschel} image. (e) and (f) Synthetic \textit{Herschel} images corresponding to (b) and (c) (see text).}
\label{fig:obs}
\end{figure*}

We first performed simulations setting the companion eccentricity to zero. According to the orbital fit, such orbits are likely to have a semi-major axis between 20 and 60~au, with a stronger probability between 20 and 30~au. On the other hand, the empirical gap-opening formula from \citet{Wisdom1980} predicts that if the cavity radius is 75~au, the companion semi-major axis has to be below 50~au. In Fig.~\ref{fig:e=0}, we represent the simulation outcome for a circular orbit of 30~au and 60~au. As predicted, the former configuration is compatible with the observations, while the latter configuration is not: the gap would be too extended. {For semi-major axes smaller than 50~au, we note that the disk would not be cleared out to the largest allowed inner cavity, which could suggest the presence of an additional companion beyond the orbit of HR~2562~B responsible for sculpting the disk. From the AMES-COND detection limits in \citet{Mesa2018}, we can exclude giant planet companions with projected separations beyond 40~au more massive than 5~$M_{\rm{J}}$ (200~Myr), 8~$M_{\rm{J}}$ (450~Myr), and 10~$M_{\rm{J}}$ (750~Myr). For projected separations beyond 60~au, the detection limits are $>$3~$M_{\rm{J}}$ (200~Myr), $>$5~$M_{\rm{J}}$ (450~Myr) and $>$6~$M_{\rm{J}}$ (750~Myr).}

We then considered eccentric orbits. Figure~\ref{fig:e} shows the simulation results for two configurations: $a$\,=\,20~au and $e$\,=\,0.8, and $a$\,=\,40~au and $e$\,=\,0.4. Because of the cavity's eccentricity, whether the outcome of a simulation matches the observations or not is not as obvious as in the circular case. As a consequence, the border between the parameter spaces of the orbital solutions compatible with the disk cavity constraints and those excluded is not well defined but is blurry. This has to be kept in mind when using the empirical gap formula {for eccentric orbits from \citet{Lazzoni2018}} to exclude orbital solutions (see top row of Fig.~\ref{fig:cornerplot}, second panel from the left).

\subsection{Comparison to \textit{Herschel} data}

Finally, we compared {the N-body images of the two simulated configurations compatible with the estimated disk cavity} to the \textit{Herschel}/PACS image at 70~$\mu$m from \citet{Moor2015}. We assumed that the population of simulated bodies is, at first order, a good tracer of the dust grains probed by \textit{Herschel}. After adequately orienting the disk plane in the simulated images, we assumed a radial temperature profile for the dust grains \citep[see Eq.~(3) in][]{Backman1993} and that the dust grains emit like black bodies. The temperatures predicted for the dust grains are $\sim$370~K at 1~au, 120~K at 10~au, and 40~K at 100~au. {For the surface density particles, we recall that it is set at the beginning of the N-body simulations and is inversely proportional to the distance to the star.} Subsequently, we used the derived temperatures to weight the contributions to the flux density of the individual particles using Planck's law. To create an image, we summed pixel by pixel all the individual contributions from particles in a column subtended by a pixel and each resulting image was convolved with the \textit{Herschel} PSF.

When comparing the synthetic \textit{Herschel} images to the measured data, we noted a large flux ratio between the inner and outer parts of the disk in the simulated images, the likes of which is not measured in the data. This feature in the simulated images appears because the inner disks, even small or with low density, are the main contributors to the disk emission. It is expected that an inner disk should be depleted in an old debris disk, such as HR~2562, because of the ``inside-out'' evolution of the dust grains \citep[e.g.,][]{Kenyon2008}. Briefly, large planetesimals will progressively disappear through collisions and the production of smaller and smaller dust grains. This evolution is must faster in the innermost regions because of the shorter dynamical timescales. This results in a large population of small grains close to the star that will be expelled from the inner disk by the stellar radiation pressure. Interestingly, \citet{Pawellek2014} found for HR~2562 a larger minimum grain size than the blowout-limit grain size, which is consistent with this scenario. {In fact, no warm disk component was identified by \citet{Moor2015} from the analysis of the target SED.} The fact that inner disks persist in our N-body simulations stems from the non-inclusion of collisions between bodies. Since the \textit{Herschel} image {and the SED of HR~2562 do not show evidence for an inner disk}, we removed the contribution from the simulated inner disks to obtain the synthetic \textit{Herschel} images shown in Fig.~\ref{fig:obs}. We find that a circular orbit and a very eccentric orbit for the companion produce similar synthetic disk images and therefore cannot be distinguished.

\begin{table}[t]
\caption{{Preliminary orbital parameters of HR~2562~B {from the combined astrometric and dynamical analysis}}.}
\label{tab:orbparams}
\begin{center}
\begin{tabular}{l c c c c c}
\hline\hline
Parameter & Unit & Median & Lower & Upper & $\chi^2_{\rm{min}}$ \\
\hline
$P$ & yr & 141 & 87 & 227 & 159 \\
$a$ & au & 30 & 22 & 41 & 32 \\
$e$ & & 0.22 & 0.07 & 0.49 & 0.19 \\
$i$ & $^{\circ}$ & 87 & 85 & 88 & 86 \\
$\Omega$ & $^{\circ}$ & 121 & 119 & 124 & 120 \\
$\omega$ & $^{\circ}$ & $-$24 & $-$88 & 82 & 143 \\
$T_0$ & & 2020 & 1934 & 2111 & 2016 \\
\hline
\end{tabular}
\end{center}
\tablefoot{The parameters are the period, semi-major axis, eccentricity, inclination, longitude of node, argument of periastron, and time at periastron.}
\end{table}

\subsection{Effects on derived orbital parameters}

We used the constraints on the companion chaotic zone and the 3-$\sigma$ estimates on the disk inclination and position angle to further refine the orbital solutions derived in the lower-left part of Fig.~\ref{fig:cornerplot} (Sect.~\ref{sec:orbit}). The strong high-eccentricity peak is strongly attenuated in the resulting distribution because of the removal of non-coplanar orbits. These additional constraints allow to sharpen the histogram distributions as shown in the upper-right part of the figure, especially for the period and the time at periastron passage. The 68\% intervals are (Table~\ref{tab:orbparams}): {$P$\,$\sim$\,87--227~yr, $e$\,$\sim$\,0.07--0.49, $i$\,$\sim$\,85--88$^{\circ}$, $\Omega$\,$\sim$\,119--124$^{\circ}$, $T_0$\,=\,1934--2111, and $a$\,$\sim$\,22--41~au (distribution not shown for the latter parameter). The distribution of arguments of periastron shows now only a marginal peak around $-$30$^{\circ}$}. The distribution of time at periastron exhibits two peaks, one sharp peak in $\sim$2000 and a broader peak in $\sim$2080.

\begin{figure}[t]
\centering
\includegraphics[trim=5mm 0mm 3mm 4mmm,clip,width=0.43\textwidth]{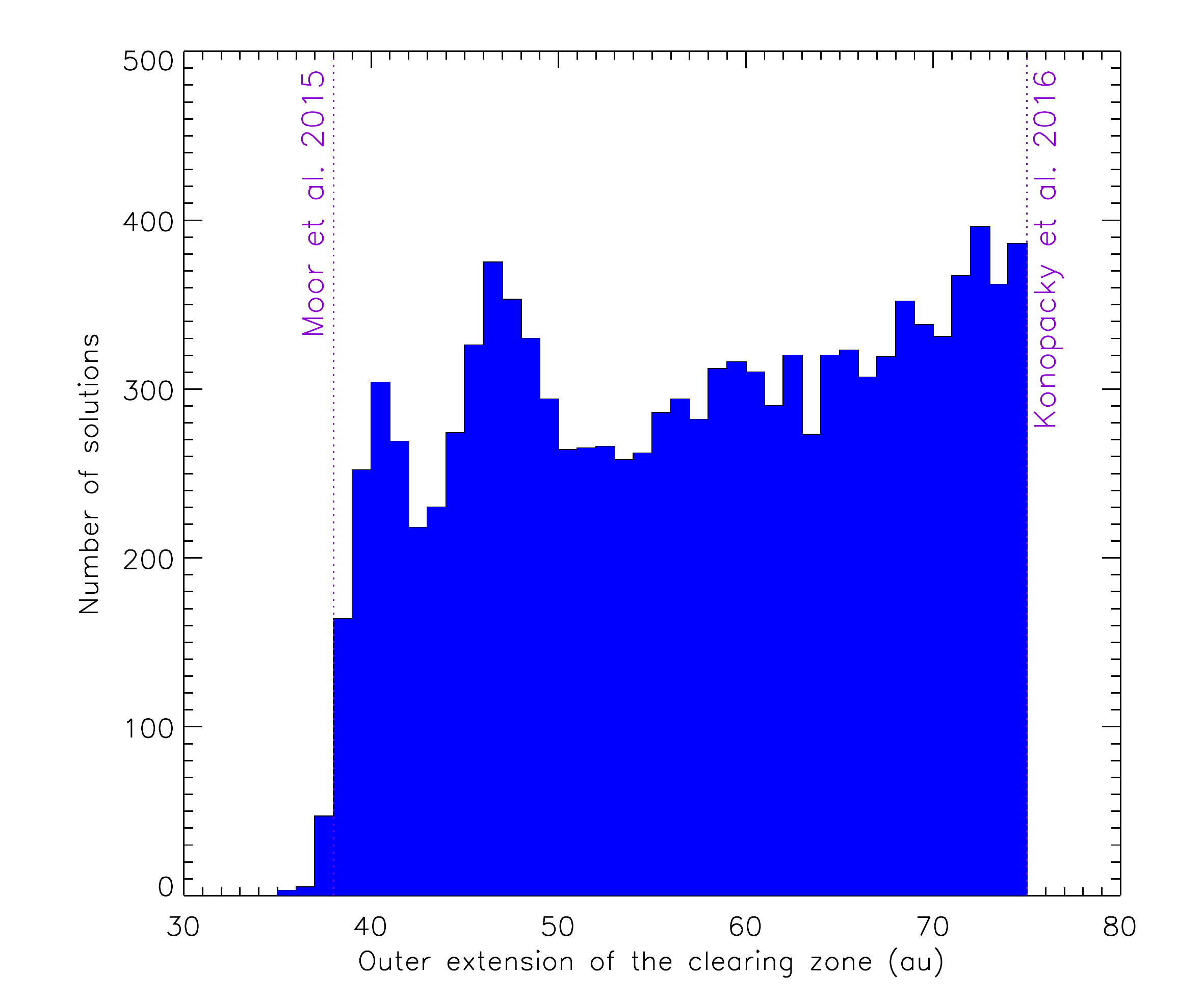}
\caption{{Distribution of the outer extent of the clearing zone of HR~2562~B for the orbital solutions compatible with the disk geometry and cavity constraints according to the relations in \citet{Lazzoni2018}. The vertical lines show the disk cavity estimates of \citet{Moor2015} and \citet{Konopacky2016b}.}}
\label{fig:aoutclear}
\end{figure}

\subsection{{Shaping of the disk cavity by HR~2562~B}}

{We finally represent in Fig.~\ref{fig:aoutclear} the distribution of the outer extent of the clearing zone associated with the orbits compatible with the disk observations assuming the relations in \citet{Lazzoni2018}. We note that the disk cavity estimate of 38~au from \citet{Moor2015} is located at the low end of the distribution. We thus conclude that the current orbit of the companion is likely responsible for the shaping of the gap. If the companion separation continues to increase in the coming years without any sign of deceleration, this would mean that the actual disk cavity edge is located further than 38~au. Depending on the outcome of further astrometric monitoring of the companion, the analysis of \citet{Moor2015} may or may not be rejected because of the large uncertainty they estimated for the inner edge of the debris belt ($\pm$20~au).}

\section{{Discussion}}

\subsection{{Formation scenarios for the companion}}

{The fact that the brown dwarf orbit is (quasi-)coplanar with the debris disk might suggest a formation process in the disk for the companion, similar to a planet-like scenario. With a mass ratio to the star of 0.02, the companion seems too massive to have formed through core accretion \citep{Mizuno1980, Pollack1996, Mordasini2012}. Its mass and semi-major axis are compatible with predictions from disk gravitational instabilities \citep{Boss1997, Forgan2013} and from collapse and fragmentation of a dense molecular cloud \citep{Bate2009}. In particular, a large companion eccentricity could be a natural outcome from a formation process by collapse with fragmentation of a dense molecular cloud, whereas it could be more difficult to explain it in a disk gravitational instability scenario (but see discussion below). In the following, we
further discuss a disk gravitational instability scenario {as a potential} formation process of HR~2562~B.}

\begin{figure}[t]
\centering
\includegraphics[trim=9mm 0mm 3mm 10mmm,clip,width=0.4\textwidth]{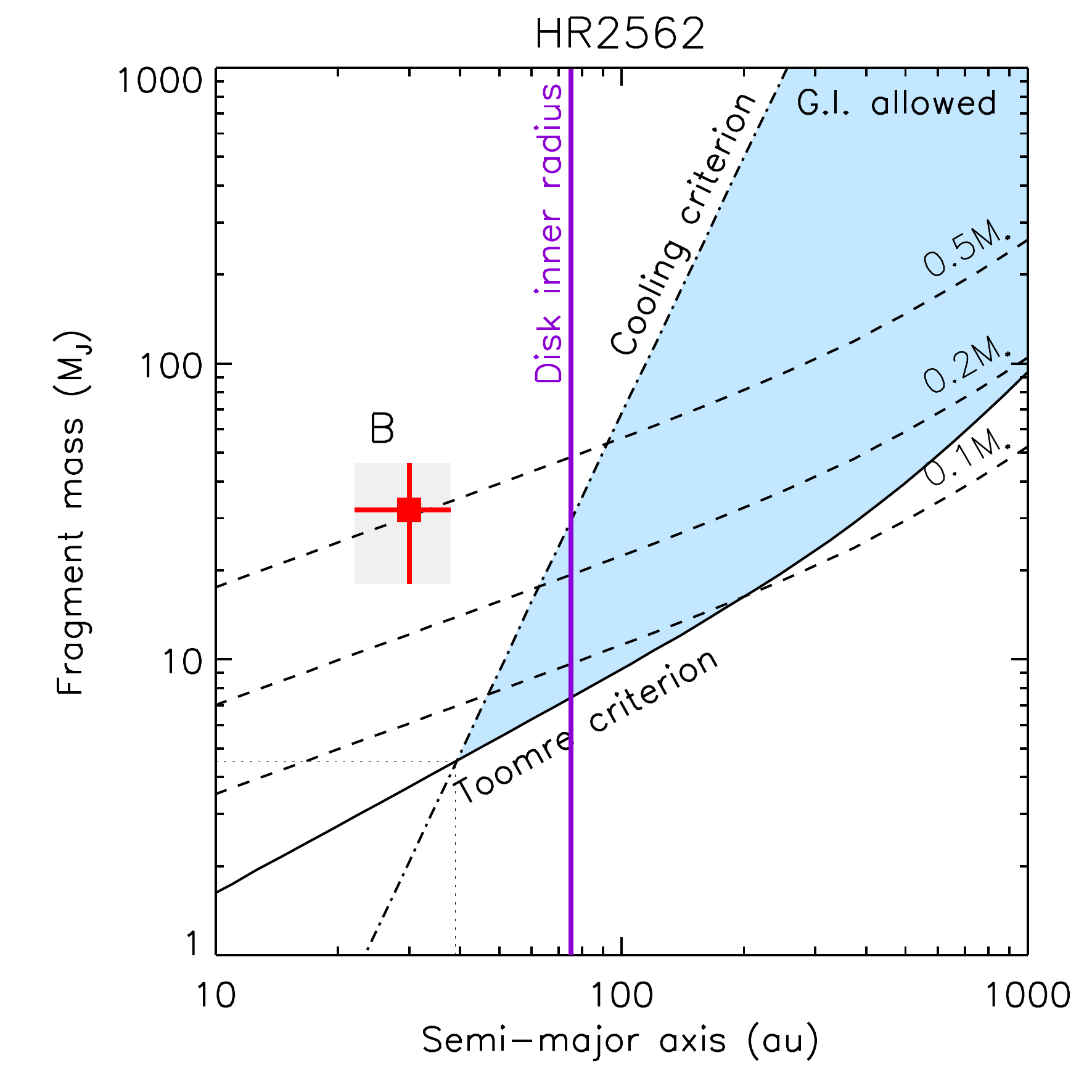}
\caption{{Masses of fragments that could be produced in-situ via disk gravitational instabilities for HR~2562 as a function of the semi-major axis to the star (blue area). Fragments with masses above the curve labeled ``Cooling criterion'' cannot cool efficiently enough, while those below the curve labeled ``Toomre criterion'' do not satisfy the Toomre criterion. The location of HR~2562~B is indicated with the red square with error bars. The primordial disk masses required to support fragments of a given mass are shown with black dashed curves for several masses expressed as fractions of the stellar mass. The purple vertical solid line indicates the inner radius of the debris disk from \citet{Konopacky2016b}.}}
\label{fig:gimodel}
\end{figure}

{We applied a disk gravitational instability model (Klahr et al., in prep.), which predicts the masses of fragments that could form in-situ following this mechanism as a function of the semi-major axis to the star. The underlying fragmentation criteria are presented in \citet{Mordasini2010} and \citet{Janson2011} and have been confirmed in local high-resolution 3D simulations \citep{Baehr2017}. Briefly, fragments can form if they satisfy the Toomre criterion for self-gravitating clumps \citep{Toomre1964} and if they are able to cool faster than the local Keplerian timescale. The model inputs include the stellar luminosity at the zero age main sequence point and the stellar metallicity. We estimated the former parameter from the isochrones of \citet{Bressan2012} and assumed for the latter parameter the value of 0.10$\pm$0.06~dex derived in \citet{Mesa2018}. The results are shown in Fig.~\ref{fig:gimodel}. We see that fragments less massive than 4~$M_{\rm{J}}$ cannot be formed at any distance to the star. HR~2562~B appears too close and too massive to have formed in-situ via disk gravitational instabilities. If we assume its nominal mass and that this mass originates from the formation process alone, it would require a very massive primordial disk with mass $\sim$40\% of the stellar mass. Such a massive primordial disk appears unlikely, because the corresponding Toomre parameter would be $<$0.2. These results combined together suggest that the companion could have formed at a larger distance to the star from a less massive fragment and subsequently migrated inward to its current location while still accreting mass from the surrounding disk material.}

{If the companion has a large eccentricity, this property might be difficult to explain in a disk gravitational instability scenario because disk interactions tend to damp the eccentricities of orbiting companions. This might imply subsequent dynamical interactions with another body to stir the eccentricity of HR~2562~B. Nevertheless, we note that for very massive substellar companions ($>$4--5~$M_{\rm{J}}$) with low inclinations to the disk plane ($<$10$^{\circ}$), numerical simulations have shown that interactions with a protoplanetary disk can stir their eccentricity \citep{Papaloizou2001, Kley2006, Bitsch2013}}.

\subsection{{Dynamical constraints on the companion mass}}

{Dynamical mass measurements of young low-mass companions offer a powerful and independent way to constrain their predicted cooling models. These models are currently highly uncertain at young ages and low masses because of the lack of observations of suitable benchmark objects. However, they are commonly used to estimate the mass of directly imaged substellar companions.}

{We used the equations in \citet{Lazzoni2018} to represent the width of the chaotic zone created by a substellar companion in units of its semi-major axis as a function of its mass ratio to HR~2562 for several orbital eccentricities in Fig.~\ref{fig:wisdommustill}. The chaotic zone width is defined as $(\Delta a/a)_{\mathrm{chaos}}$\,=\,($a_{\rm{cav}}-a$)/$a$, where $a_{\rm{cav}}$ is the cavity radius. For a given chaotic zone width and companion eccentricity, this plot gives an estimate of the companion mass. Unfortunately, these two parameters have large uncertainties so the constraints on the companion mass are quite loose and can be in the stellar regime. If we assume that the companion is a brown dwarf given the estimates on its spectral properties and that the cavity is carved exclusively by the current orbit of the companion\footnote{Other phenomena not accounted for in the formulae of \citet{Lazzoni2018} (migration of the companion, instabilities in the primordial disk, additional bodies in the system) could enlarge the cavity.}, we see that the latter cannot be on a circular orbit if the chaotic zone width is larger than $\sim$0.6. Under the same hypotheses, the chaotic zone width has to be smaller than $\sim$2.5 if the companion has an eccentricity of 0.8. For non-zero eccentricities, an even more stringent upper limit on the companion mass can be set if the eccentricity is larger and/or the chaotic zone width is smaller.  For a disk cavity radius of 38~au and a companion semi-major axis of 20~au, we see that the companion needs to have a lower eccentricity of $\sim$0.2, whereas its upper eccentricity is $\sim$0.3. For a disk cavity radius of 75~au and a companion semi-major axis of 30~au, the eccentricity constraints are $\sim$0.4--0.5. We acknowledge that these results are strongly dependent on the assumed criterion for the cavity shaping. For comparison, we show  similar diagrams to Fig.~\ref{fig:wisdommustill} in Appendix~\ref{sec:petrovich2015regaly2018} based on the equations of \citet{Petrovich2015} and \citet{Regaly2018}}.

{As already discussed in Sect.~\ref{sec:orbit}, further astrometric monitoring of the companion will be essential to measure inflexions in its orbital motion that will help to discriminate between an eccentric short-period orbit and a circular long-period orbit. On the other hand, disk observations at higher resolutions will be valuable to refine the estimates of its cavity shape and size (see Sect.~\ref{sec:almadiskobs}). Such combined constraints will provide powerful insights into the architecture of the system and the dynamical mass of the companion.}

\begin{figure}[t]
\centering
\includegraphics[trim=10mm 0mm 8mm 10mmm,width=0.4\textwidth]{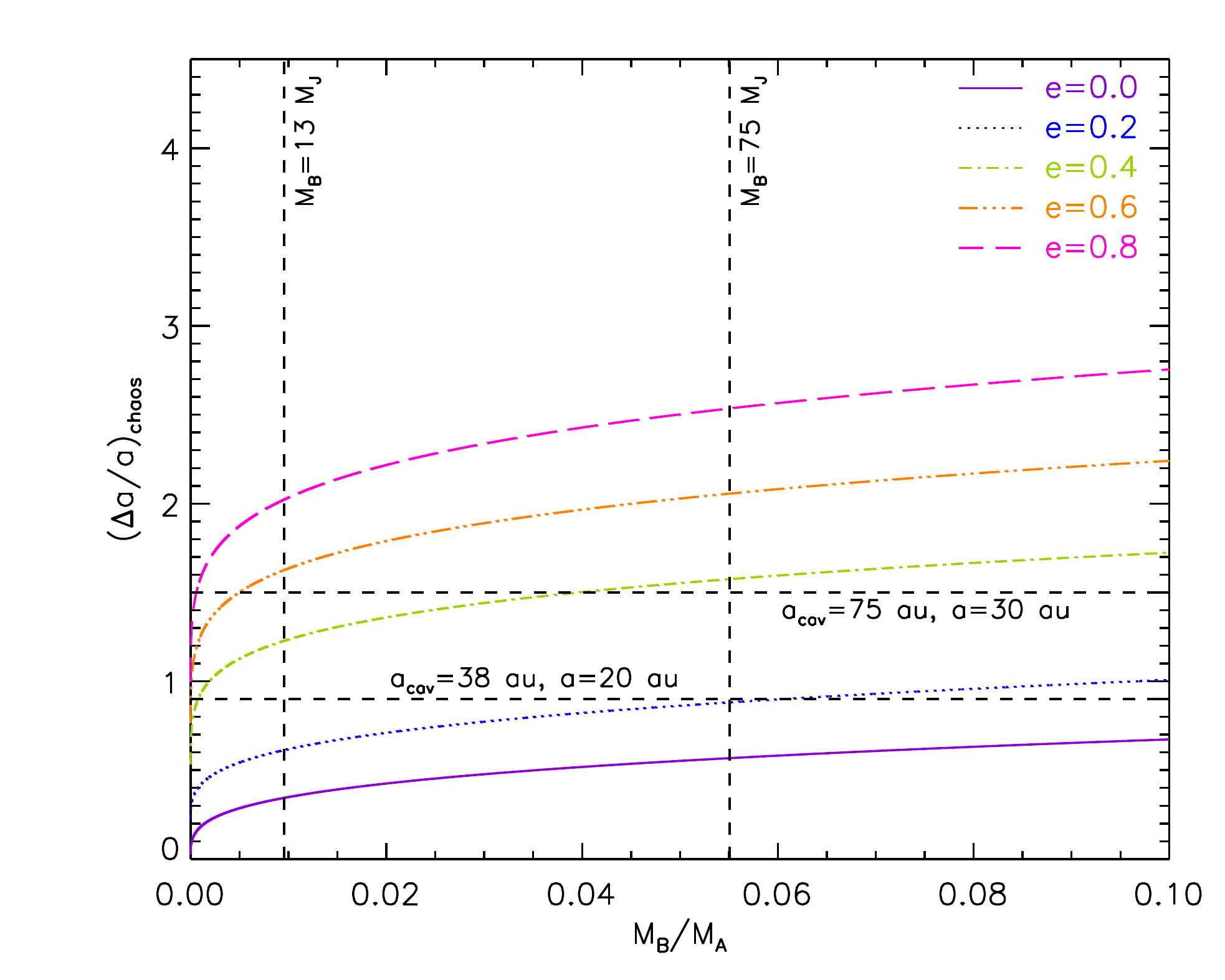}
\caption{{Width of the chaotic zone $(\Delta a/a)_{\mathrm{chaos}}$\,=\,$(a_{\rm{cav}}-a)/a$ as a function of the mass ratio to the star of a substellar companion carving the disk cavity of HR~2562 for several eccentricities according to the relations in \citet{Lazzoni2018}. The vertical dashed lines delimit the brown dwarf mass regime and the horizontal dashed lines show examples of chaotic zone width for two (disk cavity radius, companion semi-major axis) couples.}}
\label{fig:wisdommustill}
\end{figure}

\subsection{{Constraints on the disk properties from ALMA data}}
\label{sec:almadiskobs}

{(Sub-)millimeter observations of the HR~2562 disk at high angular resolutions with ALMA will be valuable to further refine the estimates on the extent and shape of its cavity. \citet{Regaly2018} discuss the potential of ALMA data for analyzing dynamical interactions between substellar companions and debris disks. In particular, they provide a method to estimate the orbital eccentricity and mass of a giant planet carving the disk cavity. This involves measurements of the cavity size and offset with respect to the star by ellipse fitting to a given intensity contour level, which itself depends on the image resolution (the optimal contour level is larger for poorer resolutions). The robustness of the empirical relations was checked against the planetesimals' initial eccentricity and inclination by simulating hot and cold disks and against the stellar mass and age for ranges of 0.6--1~Gyr and 0.5--2~$M_{\odot}$, respectively. They found relations for the cavity size and for the cavity center offset with respect to the star which only depend on the planet/star mass ratio and the planet eccentricity for eccentric orbits, allowing to break the degeneracies between these two unknowns. They also show that the cavity eccentricity cannot be used as a direct proxy for the planet eccentricity because they are not identical and their relation is not a monotonic function. A disk cavity can be eccentric while a perturbing planet orbit is circular. The eccentricity of the disk cavity is only equal to that of the giant planet perturber for a narrow range of intermediate planet eccentricities (0.3--0.6 for a 5-$M_{\rm{J}}$ giant planet). Another observable disk feature that could be suggestive of a large planetary eccentricity outlined by \citet{Regaly2018} would be the detection of an azimuthal brightness asymmetry or ``glow'' with a large contrast (up to $\sim$50\% for a 5-$M_{\rm{J}}$ planet) located beyond the disk cavity wall and near the position angle of the planet apastron.}

{Using the relations for the cavity size and offset with respect to the star in \citet{Regaly2018}, we estimated that if these quantities could be measured with accuracies of $\sim$10\%\footnote{{Limitations to the accuracy of these measurements include the shape of the instrument beam (an elliptical beam can introduce artifacts in the images like brightness asymmetries), instrument pointing accuracy, and the scatter induced by the planet orbital phase.}}, the eccentricity and mass ratio to the star of the planet could be assessed with accuracies of $\sim$20\% and $\sim$40\%, respectively. The stellar mass being constrained with an accuracy of $<$2\% \citep{Mesa2018}, the dynamical mass estimate of HR~2562~B would be slightly more accurate than the $\sim$45--50\% uncertainties of the evolutionary model predictions \citep{Konopacky2016b, Mesa2018} but independent from assumptions on the formation mechanism and the system age.}

{Such disk cavity measurements require high-resolution images. With a diameter of 0.5$''$ for the ALMA instrument beam, {the HR~2562 disk cavity would be resolved with $\sim$4.5 resolution elements for the smallest diameter estimate of 38$\times$2~au from \citet{Moor2015}, and $\sim$9 resolution elements for the largest diameter estimate of 75$\times$2~au from \citet{Konopacky2016b}}. These resolutions are in the range of the resolutions for which the methods proposed by \citet{Regaly2018} could be applied. Higher resolutions could be achieved but at the cost of longer integration times to compensate for the instrument sensitivity loss.}

\section{Summary}

We present VLT/SPHERE observations of the young system of HR~2562 to redetect and further characterize the orbit of its brown dwarf companion. The SPHERE data show a strong increase of the companion separation of $\sim$40~mas ($\sim$1.3~au) over 1.7~yr with respect to the GPI measurements, ruling out a face-on circular orbit. The joint fit of the SPHERE and GPI astrometry clearly indicates for the companion an orbit (quasi-)coplanar with the known debris disk without any prior on the orbital plane. Furthermore, the eccentricity distribution suggests a non-zero eccentricity, which could reconcile the mass estimates from the evolutionary models and from dynamical considerations assuming that the object is responsible for the truncation of the debris belt. Assuming a debris belt inner edge at 75~au, a dynamical analysis based on analytical and numerical approaches allows to reject eccentricities larger than $\sim$0.3 for periods longer than 200~yr and eccentricities smaller than 0.15 for periods shorter than 100~yr. {If the companion has formed through disk gravitational instabilities, our analysis suggests that its current location and mass can be accounted for by formation at a larger distance to the star from a less massive disk fragment followed by inward migration with mass accretion.} Further astrometric monitoring of the companion in order to detect curvature in its orbital motion will allow to better constrain its period and eccentricity. In addition, far-IR or millimeter images at higher resolutions are needed to {determine more precisely the disk geometry and its cavity extent}. With such information combined with a lower limit on the orbital eccentricity of the companion, the dynamical mass of HR~2562~B could be strongly constrained, making it a valuable benchmark object close to the L/T transition for evolutionary and atmospheric models of substellar companions.

\begin{acknowledgements}
     The authors thank the ESO Paranal Staff for support in conducting the observations and Philippe Delorme and Eric Lagadec (SPHERE Data Center) for their help with the data reduction. We also thank Attila Mo\'or for kindly providing the \textit{Herschel} data, Grant Kennedy, Virginie Faramaz, and Nicole Pawellek for useful discussions about the system, and Grant Kennedy for his public imorbel web interface. {We finally thank the referee, Alexander Mustill, for a constructive report that helped to improve the manuscript.} We acknowledge financial support from the Programme National de Planétologie (PNP) and the Programme National de Physique Stellaire (PNPS) of CNRS-INSU. This work has also been supported by a grant from the French LabEx OSUG@2020 (Investissements d'avenir -- ANR10 LABX56). The project is supported by CNRS, by the Agence Nationale de la Recherche (ANR-14-CE33-0018). J.\,O. acknowledges financial support from ICM N\'ucleo Milenio de Formaci\'on Planetaria, NPF. This work has made use of the SPHERE Data Centre, jointly operated by OSUG/IPAG (Grenoble), PYTHEAS/LAM/CeSAM (Marseille), OCA/Lagrange (Nice) and Observatoire de Paris/LESIA (Paris). This research made use of the SIMBAD database and the VizieR Catalogue access tool, both operated at the CDS, Strasbourg, France. The original description of the VizieR service was published in Ochsenbein et al. (2000, A\&AS 143, 23). This research has made use of NASA's Astrophysics Data System Bibliographic Services. SPHERE is an instrument designed and built by a consortium consisting of IPAG (Grenoble, France), MPIA (Heidelberg, Germany), LAM (Marseille, France), LESIA (Paris, France), Laboratoire Lagrange (Nice, France), INAF -- Osservatorio di Padova (Italy), Observatoire astronomique de l'Universit\'e de Gen\`eve (Switzerland), ETH Zurich (Switzerland), NOVA (Netherlands), ONERA (France), and ASTRON (Netherlands), in collaboration with ESO. SPHERE was funded by ESO, with additional contributions from the CNRS (France), MPIA (Germany), INAF (Italy), FINES (Switzerland), and NOVA (Netherlands). SPHERE also received funding from the European Commission Sixth and Seventh Framework Programs as part of the Optical Infrared Coordination Network for Astronomy (OPTICON) under grant number RII3-Ct-2004-001566 for FP6 (2004--2008), grant number 226604 for FP7 (2009--2012), and grant number 312430 for FP7 (2013--2016).

\end{acknowledgements}

%
   \bibliographystyle{aa} 
   \bibliography{biblio} 

\begin{thebibliography}{77}
\expandafter\ifx\csname natexlab\endcsname\relax\def\natexlab#1{#1}\fi

\bibitem[{{Asiain} {et~al.}(1999){Asiain}, {Figueras}, {Torra}, \&
  {Chen}}]{Asiain1999}
{Asiain}, R., {Figueras}, F., {Torra}, J., \& {Chen}, B. 1999, \aap, 341, 427

\bibitem[{{Backman} \& {Paresce}(1993)}]{Backman1993}
{Backman}, D.~E. \& {Paresce}, F. 1993, in Protostars and Planets III, ed.
  E.~H. {Levy} \& J.~I. {Lunine}, 1253--1304

\bibitem[{{Baehr} {et~al.}(2017){Baehr}, {Klahr}, \& {Kratter}}]{Baehr2017}
{Baehr}, H., {Klahr}, H., \& {Kratter}, K.~M. 2017, \apj, 848, 40

\bibitem[{{Baraffe} {et~al.}(2003){Baraffe}, {Chabrier}, {Barman}, {Allard}, \&
  {Hauschildt}}]{Baraffe2003}
{Baraffe}, I., {Chabrier}, G., {Barman}, T.~S., {Allard}, F., \& {Hauschildt},
  P.~H. 2003, A\&A, 402, 701

\bibitem[{{Baraffe} {et~al.}(2015){Baraffe}, {Homeier}, {Allard}, \&
  {Chabrier}}]{Baraffe2015}
{Baraffe}, I., {Homeier}, D., {Allard}, F., \& {Chabrier}, G. 2015, A\&A, 577,
  A42

\bibitem[{{Bate}(2009)}]{Bate2009}
{Bate}, M.~R. 2009, MNRAS, 392, 590

\bibitem[{{Beuzit} {et~al.}(2008){Beuzit}, {Feldt}, {Dohlen}, {Mouillet},
  {Puget}, {Wildi}, {Abe}, {Antichi}, {Baruffolo}, {Baudoz}, {Boccaletti},
  {Carbillet}, {Charton}, {Claudi}, {Downing}, {Fabron}, {Feautrier},
  {Fedrigo}, {Fusco}, {Gach}, {Gratton}, {Henning}, {Hubin}, {Joos}, {Kasper},
  {Langlois}, {Lenzen}, {Moutou}, {Pavlov}, {Petit}, {Pragt}, {Rabou}, {Rigal},
  {Roelfsema}, {Rousset}, {Saisse}, {Schmid}, {Stadler}, {Thalmann}, {Turatto},
  {Udry}, {Vakili}, \& {Waters}}]{Beuzit2008}
{Beuzit}, J.-L., {Feldt}, M., {Dohlen}, K., {et~al.} 2008, in SPIE Conf. Ser.,
  ed. I.~S. {McLean} \& M.~M. {Casali}, Vol. 7014, 701418

\bibitem[{{Bitsch} {et~al.}(2013){Bitsch}, {Crida}, {Libert}, \&
  {Lega}}]{Bitsch2013}
{Bitsch}, B., {Crida}, A., {Libert}, A.-S., \& {Lega}, E. 2013, \aap, 555, A124

\bibitem[{{Boccaletti} {et~al.}(2018){Boccaletti}, {Sezestre}, {Lagrange},
  {Th{\'e}bault}, {Gratton}, {Langlois}, {Thalmann}, {Janson}, {Delorme},
  {Augereau}, {Schneider}, {Milli}, {Grady}, {Debes}, {Kral}, {Olofsson},
  {Carson}, {Maire}, {Henning}, {Wisniewski}, {Schlieder}, {Dominik},
  {Desidera}, {Ginski}, {Hines}, {M{\'e}nard}, {Mouillet}, {Pawellek}, {Vigan},
  {Lagadec}, {Avenhaus}, {Beuzit}, {Biller}, {Bonavita}, {Bonnefoy},
  {Brandner}, {Cantalloube}, {Chauvin}, {Cheetham}, {Cudel}, {Gry}, {Daemgen},
  {Feldt}, {Galicher}, {Girard}, {Hagelberg}, {Janin-Potiron}, {Kasper},
  {Coroller}, {Mesa}, {Peretti}, {Perrot}, {Samland}, {Sissa}, {Wildi},
  {Zurlo}, {Rochat}, {Stadler}, {Gluck}, {Orign{\'e}}, {Llored}, {Baudoz},
  {Rousset}, {Martinez}, \& {Rigal}}]{Boccaletti2018}
{Boccaletti}, A., {Sezestre}, E., {Lagrange}, A.-M., {et~al.} 2018, \aap, 614,
  A52

\bibitem[{{Boss}(1997)}]{Boss1997}
{Boss}, A.~P. 1997, Science, 276, 1836

\bibitem[{{Bressan} {et~al.}(2012){Bressan}, {Marigo}, {Girardi}, {Salasnich},
  {Dal Cero}, {Rubele}, \& {Nanni}}]{Bressan2012}
{Bressan}, A., {Marigo}, P., {Girardi}, L., {et~al.} 2012, \mnras, 427, 127

\bibitem[{{Cantalloube} {et~al.}(2015){Cantalloube}, {Mouillet}, {Mugnier},
  {Milli}, {Absil}, {Gomez Gonzalez}, {Chauvin}, {Beuzit}, \&
  {Cornia}}]{Cantalloube2015}
{Cantalloube}, F., {Mouillet}, D., {Mugnier}, L.~M., {et~al.} 2015, \aap, 582,
  A89

\bibitem[{{Carbillet} {et~al.}(2011){Carbillet}, {Bendjoya}, {Abe}, {Guerri},
  {Boccaletti}, {Daban}, {Dohlen}, {Ferrari}, {Robbe-Dubois}, {Douet}, \&
  {Vakili}}]{Carbillet2011}
{Carbillet}, M., {Bendjoya}, P., {Abe}, L., {et~al.} 2011, Experimental
  Astronomy, 30, 39

\bibitem[{{Casagrande} {et~al.}(2011){Casagrande}, {Sch{\"o}nrich}, {Asplund},
  {Cassisi}, {Ram{\'{\i}}rez}, {Mel{\'e}ndez}, {Bensby}, \&
  {Feltzing}}]{Casagrande2011}
{Casagrande}, L., {Sch{\"o}nrich}, R., {Asplund}, M., {et~al.} 2011, \aap, 530,
  A138

\bibitem[{{Chauvin} {et~al.}(2017){Chauvin}, {Desidera}, {Lagrange}, {Vigan},
  {Feldt}, {Gratton}, {Langlois}, {Cheetham}, {Bonnefoy}, \&
  {Meyer}}]{Chauvin2017b}
{Chauvin}, G., {Desidera}, S., {Lagrange}, A.-M., {et~al.} 2017, in SF2A-2017:
  Proceedings of the Annual meeting of the French Society of Astronomy and
  Astrophysics, ed. C.~{Reyl{\'e}}, P.~{Di Matteo}, F.~{Herpin}, E.~{Lagadec},
  A.~{Lan{\c c}on}, Z.~{Meliani}, \& F.~{Royer}, 331--335

\bibitem[{{Chauvin} {et~al.}(2012){Chauvin}, {Lagrange}, {Beust}, {Bonnefoy},
  {Boccaletti}, {Apai}, {Allard}, {Ehrenreich}, {Girard}, {Mouillet}, \&
  {Rouan}}]{Chauvin2012}
{Chauvin}, G., {Lagrange}, A.-M., {Beust}, H., {et~al.} 2012, A\&A, 542, A41

\bibitem[{{Claudi} {et~al.}(2008){Claudi}, {Turatto}, {Gratton}, {Antichi},
  {Bonavita}, {Bruno}, {Cascone}, {De Caprio}, {Desidera}, {Giro}, {Mesa},
  {Scuderi}, {Dohlen}, {Beuzit}, \& {Puget}}]{Claudi2008}
{Claudi}, R.~U., {Turatto}, M., {Gratton}, R.~G., {et~al.} 2008, in SPIE Conf.
  Ser., Vol. 7014, 70143E

\bibitem[{{Delorme} {et~al.}(2017){Delorme}, {Meunier}, {Albert}, {Lagadec},
  {Le Coroller}, {Galicher}, {Mouillet}, {Boccaletti}, {Mesa}, {Meunier},
  {Beuzit}, {Lagrange}, {Chauvin}, {Sapone}, {Langlois}, {Maire},
  {Montarg{\`e}s}, {Gratton}, {Vigan}, \& {Surace}}]{Delorme2017b}
{Delorme}, P., {Meunier}, N., {Albert}, D., {et~al.} 2017, in SF2A-2017:
  Proceedings of the Annual meeting of the French Society of Astronomy and
  Astrophysics, ed. C.~{Reyl{\'e}}, P.~{Di Matteo}, F.~{Herpin}, E.~{Lagadec},
  A.~{Lan{\c c}on}, Z.~{Meliani}, \& F.~{Royer}, 347--361

\bibitem[{{Desidera} {et~al.}(2015){Desidera}, {Covino}, {Messina}, {Carson},
  {Hagelberg}, {Schlieder}, {Biazzo}, {Alcal{\'a}}, {Chauvin}, {Vigan},
  {Beuzit}, {Bonavita}, {Bonnefoy}, {Delorme}, {D'Orazi}, {Esposito}, {Feldt},
  {Girardi}, {Gratton}, {Henning}, {Lagrange}, {Lanzafame}, {Launhardt},
  {Marmier}, {Melo}, {Meyer}, {Mouillet}, {Moutou}, {Segransan}, {Udry}, \&
  {Zaidi}}]{Desidera2015}
{Desidera}, S., {Covino}, E., {Messina}, S., {et~al.} 2015, \aap, 573, A126

\bibitem[{{Dohlen} {et~al.}(2008){Dohlen}, {Langlois}, {Saisse}, {Hill},
  {Origne}, {Jacquet}, {Fabron}, {Blanc}, {Llored}, {Carle}, {Moutou}, {Vigan},
  {Boccaletti}, {Carbillet}, {Mouillet}, \& {Beuzit}}]{Dohlen2008a}
{Dohlen}, K., {Langlois}, M., {Saisse}, M., {et~al.} 2008, in SPIE Conf. Ser.,
  Vol. 7014, 70143L

\bibitem[{{Dohnanyi}(1969)}]{Dohnanyi1969}
{Dohnanyi}, J.~S. 1969, \jgr, 74, 2531

\bibitem[{{Eastman} {et~al.}(2013){Eastman}, {Gaudi}, \& {Agol}}]{Eastman2013}
{Eastman}, J., {Gaudi}, B.~S., \& {Agol}, E. 2013, \pasp, 125, 83

\bibitem[{{Esposito} {et~al.}(2013){Esposito}, {Mesa}, {Skemer}, {Arcidiacono},
  {Claudi}, {Desidera}, {Gratton}, {Mannucci}, {Marzari}, {Masciadri}, {Close},
  {Hinz}, {Kulesa}, {McCarthy}, {Males}, {Agapito}, {Argomedo}, {Boutsia},
  {Briguglio}, {Brusa}, {Busoni}, {Cresci}, {Fini}, {Fontana}, {Guerra},
  {Hill}, {Miller}, {Paris}, {Pinna}, {Puglisi}, {Quiros-Pacheco}, {Riccardi},
  {Stefanini}, {Testa}, {Xompero}, \& {Woodward}}]{Esposito2013}
{Esposito}, S., {Mesa}, D., {Skemer}, A., {et~al.} 2013, A\&A, 549, A52

\bibitem[{{Forgan} \& {Rice}(2013)}]{Forgan2013}
{Forgan}, D. \& {Rice}, K. 2013, \mnras, 432, 3168

\bibitem[{{Gaia Collaboration} {et~al.}(2016){Gaia Collaboration}, {Brown},
  {Vallenari}, {Prusti}, {de Bruijne}, {Mignard}, {Drimmel}, {Babusiaux},
  {Bailer-Jones}, {Bastian}, \& et~al.}]{GaiaCollaboration2016}
{Gaia Collaboration}, {Brown}, A.~G.~A., {Vallenari}, A., {et~al.} 2016, A\&A,
  595, A2

\bibitem[{{Galicher} {et~al.}(2018){Galicher}, {Boccaletti}, {Mesa}, {Delorme},
  {Gratton}, {Langlois}, {Lagrange}, {Maire}, {Le Coroller}, {Chauvin},
  {Biller}, {Cantalloube}, {Janson}, {Lagadec}, {Meunier}, {Vigan},
  {Hagelberg}, {Bonnefoy}, {Zurlo}, {Rocha}, {Maurel}, {Jaquet}, {Buey}, \&
  {Weber}}]{Galicher2018}
{Galicher}, R., {Boccaletti}, A., {Mesa}, D., {et~al.} 2018, \aap, 615, A92

\bibitem[{{Galicher} \& {Marois}(2011)}]{Galicher2011a}
{Galicher}, R. \& {Marois}, C. 2011, in Proc. AO4ELT2, id.P25

\bibitem[{{Gray} {et~al.}(2006){Gray}, {Corbally}, {Garrison}, {McFadden},
  {Bubar}, {McGahee}, {O'Donoghue}, \& {Knox}}]{Gray2006}
{Gray}, R.~O., {Corbally}, C.~J., {Garrison}, R.~F., {et~al.} 2006, \aj, 132,
  161

\bibitem[{{Holman} \& {Wiegert}(1999)}]{Holman1999}
{Holman}, M.~J. \& {Wiegert}, P.~A. 1999, \aj, 117, 621

\bibitem[{{Janson} {et~al.}(2011){Janson}, {Bonavita}, {Klahr},
  {Lafreni{\`e}re}, {Jayawardhana}, \& {Zinnecker}}]{Janson2011}
{Janson}, M., {Bonavita}, M., {Klahr}, H., {et~al.} 2011, \apj, 736, 89

\bibitem[{{Kenyon} \& {Bromley}(2008)}]{Kenyon2008}
{Kenyon}, S.~J. \& {Bromley}, B.~C. 2008, \apjs, 179, 451

\bibitem[{{Khorrami} {et~al.}(2016){Khorrami}, {Lanz}, {Vakili}, {Lagadec},
  {Langlois}, {Brandner}, {Chesneau}, {Meyer}, {Carbillet}, {Abe}, {Mouillet},
  {Beuzit}, {Boccaletti}, {Perrot}, {Thalmann}, {Schmid}, {Pavlov}, {Costille},
  {Dohlen}, {Le Mignant}, {Petit}, \& {Sauvage}}]{Khorrami2016}
{Khorrami}, Z., {Lanz}, T., {Vakili}, F., {et~al.} 2016, A\&A, 588, L7

\bibitem[{{Kley} \& {Dirksen}(2006)}]{Kley2006}
{Kley}, W. \& {Dirksen}, G. 2006, \aap, 447, 369

\bibitem[{{Konopacky} {et~al.}(2016{\natexlab{a}}){Konopacky}, {Marois},
  {Macintosh}, {Galicher}, {Barman}, {Metchev}, \&
  {Zuckerman}}]{Konopacky2016a}
{Konopacky}, Q.~M., {Marois}, C., {Macintosh}, B.~A., {et~al.}
  2016{\natexlab{a}}, AJ, 152, 28

\bibitem[{{Konopacky} {et~al.}(2016{\natexlab{b}}){Konopacky}, {Rameau},
  {Duch{\^e}ne}, {Filippazzo}, {Giorla Godfrey}, {Marois}, {Nielsen}, {Pueyo},
  {Rafikov}, {Rice}, {Wang}, {Ammons}, {Bailey}, {Barman}, {Bulger},
  {Bruzzone}, {Chilcote}, {Cotten}, {Dawson}, {De Rosa}, {Doyon}, {Esposito},
  {Fitzgerald}, {Follette}, {Goodsell}, {Graham}, {Greenbaum}, {Hibon}, {Hung},
  {Ingraham}, {Kalas}, {Lafreni{\`e}re}, {Larkin}, {Macintosh}, {Maire},
  {Marchis}, {Marley}, {Matthews}, {Metchev}, {Millar-Blanchaer},
  {Oppenheimer}, {Palmer}, {Patience}, {Perrin}, {Poyneer}, {Rajan},
  {Rantakyr{\"o}}, {Savransky}, {Schneider}, {Sivaramakrishnan}, {Song},
  {Soummer}, {Thomas}, {Wallace}, {Ward-Duong}, {Wiktorowicz}, \&
  {Wolff}}]{Konopacky2016b}
{Konopacky}, Q.~M., {Rameau}, J., {Duch{\^e}ne}, G., {et~al.}
  2016{\natexlab{b}}, ApJL, 829, L4

\bibitem[{{Kral} {et~al.}(2017){Kral}, {Matr{\`a}}, {Wyatt}, \&
  {Kennedy}}]{Kral2017}
{Kral}, Q., {Matr{\`a}}, L., {Wyatt}, M.~C., \& {Kennedy}, G.~M. 2017, \mnras,
  469, 521

\bibitem[{{Krivov} {et~al.}(2013){Krivov}, {Eiroa}, {L{\"o}hne}, {Marshall},
  {Montesinos}, {del Burgo}, {Absil}, {Ardila}, {Augereau}, {Bayo}, {Bryden},
  {Danchi}, {Ertel}, {Lebreton}, {Liseau}, {Mora}, {Mustill}, {Mutschke},
  {Neuh{\"a}user}, {Pilbratt}, {Roberge}, {Schmidt}, {Stapelfeldt},
  {Th{\'e}bault}, {Vitense}, {White}, \& {Wolf}}]{Krivov2013}
{Krivov}, A.~V., {Eiroa}, C., {L{\"o}hne}, T., {et~al.} 2013, \apj, 772, 32

\bibitem[{{Lagrange} {et~al.}(2010){Lagrange}, {Bonnefoy}, {Chauvin}, {Apai},
  {Ehrenreich}, {Boccaletti}, {Gratadour}, {Rouan}, {Mouillet}, {Lacour}, \&
  {Kasper}}]{Lagrange2010b}
{Lagrange}, A.-M., {Bonnefoy}, M., {Chauvin}, G., {et~al.} 2010, Science, 329,
  57

\bibitem[{{Langlois} {et~al.}(2013){Langlois}, {Vigan}, {Moutou}, {Sauvage},
  {Dohlen}, {Costille}, {Mouillet}, \& {Le Mignant}}]{Langlois2013}
{Langlois}, M., {Vigan}, A., {Moutou}, C., {et~al.} 2013, in Proceedings of the
  Third AO4ELT Conference, ed. S.~{Esposito} \& L.~{Fini}, 63

\bibitem[{{Lazzoni} {et~al.}(2018){Lazzoni}, {Desidera}, {Marzari},
  {Boccaletti}, {Langlois}, {Mesa}, {Gratton}, {Kral}, {Pawellek}, {Olofsson},
  {Bonnefoy}, {Chauvin}, {Lagrange}, {Vigan}, {Sissa}, {Antichi}, {Avenhaus},
  {Baruffolo}, {Baudino}, {Bazzon}, {Beuzit}, {Biller}, {Bonavita}, {Brandner},
  {Bruno}, {Buenzli}, {Cantalloube}, {Cascone}, {Cheetham}, {Claudi}, {Cudel},
  {Daemgen}, {De Caprio}, {Delorme}, {Fantinel}, {Farisato}, {Feldt},
  {Galicher}, {Ginski}, {Girard}, {Giro}, {Janson}, {Hagelberg}, {Henning},
  {Incorvaia}, {Kasper}, {Kopytova}, {LeCoroller}, {Lessio}, {Ligi}, {Maire},
  {M{\'e}nard}, {Meyer}, {Milli}, {Mouillet}, {Peretti}, {Perrot}, {Rouan},
  {Samland}, {Salasnich}, {Salter}, {Schmidt}, {Scuderi}, {Sezestre},
  {Turatto}, {Udry}, {Wildi}, \& {Zurlo}}]{Lazzoni2018}
{Lazzoni}, C., {Desidera}, S., {Marzari}, F., {et~al.} 2018, \aap, 611, A43

\bibitem[{{Levison} \& {Duncan}(1994)}]{Levison1994}
{Levison}, H.~F. \& {Duncan}, M.~J. 1994, \icarus, 108, 18

\bibitem[{{Macintosh} {et~al.}(2015){Macintosh}, {Graham}, {Barman}, {De Rosa},
  {Konopacky}, {Marley}, {Marois}, {Nielsen}, {Pueyo}, {Rajan}, {Rameau},
  {Saumon}, {Wang}, {Patience}, {Ammons}, {Arriaga}, {Artigau}, {Beckwith},
  {Brewster}, {Bruzzone}, {Bulger}, {Burningham}, {Burrows}, {Chen}, {Chiang},
  {Chilcote}, {Dawson}, {Dong}, {Doyon}, {Draper}, {Duch{\^e}ne}, {Esposito},
  {Fabrycky}, {Fitzgerald}, {Follette}, {Fortney}, {Gerard}, {Goodsell},
  {Greenbaum}, {Hibon}, {Hinkley}, {Cotten}, {Hung}, {Ingraham},
  {Johnson-Groh}, {Kalas}, {Lafreniere}, {Larkin}, {Lee}, {Line}, {Long},
  {Maire}, {Marchis}, {Matthews}, {Max}, {Metchev}, {Millar-Blanchaer},
  {Mittal}, {Morley}, {Morzinski}, {Murray-Clay}, {Oppenheimer}, {Palmer},
  {Patel}, {Perrin}, {Poyneer}, {Rafikov}, {Rantakyr{\"o}}, {Rice}, {Rojo},
  {Rudy}, {Ruffio}, {Ruiz}, {Sadakuni}, {Saddlemyer}, {Salama}, {Savransky},
  {Schneider}, {Sivaramakrishnan}, {Song}, {Soummer}, {Thomas}, {Vasisht},
  {Wallace}, {Ward-Duong}, {Wiktorowicz}, {Wolff}, \&
  {Zuckerman}}]{Macintosh2015}
{Macintosh}, B., {Graham}, J.~R., {Barman}, T., {et~al.} 2015, Science, 350, 64

\bibitem[{{Maire} {et~al.}(2016){Maire}, {Langlois}, {Dohlen}, {Lagrange},
  {Gratton}, {Chauvin}, {Desidera}, {Girard}, {Milli}, {Vigan}, {Zins},
  {Delorme}, {Beuzit}, {Claudi}, {Feldt}, {Mouillet}, {Puget}, {Turatto}, \&
  {Wildi}}]{Maire2016b}
{Maire}, A.-L., {Langlois}, M., {Dohlen}, K., {et~al.} 2016, in SPIE Conf.
  Ser., Vol. 9908, 990834

\bibitem[{{Maire} {et~al.}(2015){Maire}, {Skemer}, {Hinz}, {Desidera},
  {Esposito}, {Gratton}, {Marzari}, {Skrutskie}, {Biller}, {Defr{\`e}re},
  {Bailey}, {Leisenring}, {Apai}, {Bonnefoy}, {Brandner}, {Buenzli}, {Claudi},
  {Close}, {Crepp}, {De Rosa}, {Eisner}, {Fortney}, {Henning}, {Hofmann},
  {Kopytova}, {Males}, {Mesa}, {Morzinski}, {Oza}, {Patience}, {Pinna},
  {Rajan}, {Schertl}, {Schlieder}, {Su}, {Vaz}, {Ward-Duong}, {Weigelt}, \&
  {Woodward}}]{Maire2015}
{Maire}, A.-L., {Skemer}, A.~J., {Hinz}, P.~M., {et~al.} 2015, A\&A, 576, A133

\bibitem[{{Marois} {et~al.}(2014){Marois}, {Correia}, {Galicher}, {Ingraham},
  {Macintosh}, {Currie}, \& {De Rosa}}]{Marois2014}
{Marois}, C., {Correia}, C., {Galicher}, R., {et~al.} 2014, in SPIE Conf. Ser.,
  Vol. 9148, 91480U

\bibitem[{Marois {et~al.}(2008)Marois, Macintosh, Barman, Zuckerman, Song,
  Patience, Lafreni{\`e}re, \& Doyon}]{Marois2008c}
Marois, C., Macintosh, B., Barman, T., {et~al.} 2008, Science, 322, 1348

\bibitem[{{Marois} {et~al.}(2010){Marois}, {Zuckerman}, {Konopacky},
  {Macintosh}, \& {Barman}}]{Marois2010b}
{Marois}, C., {Zuckerman}, B., {Konopacky}, Q., {Macintosh}, B., \& {Barman},
  T. 2010, Nature, 468, 1080

\bibitem[{{Martinez} {et~al.}(2009){Martinez}, {Dorrer}, {Aller Carpentier},
  {Kasper}, {Boccaletti}, {Dohlen}, \& {Yaitskova}}]{Martinez2009}
{Martinez}, P., {Dorrer}, C., {Aller Carpentier}, E., {et~al.} 2009, \aap, 495,
  363

\bibitem[{{Mawet} {et~al.}(2014){Mawet}, {Milli}, {Wahhaj}, {Pelat}, {Absil},
  {Delacroix}, {Boccaletti}, {Kasper}, {Kenworthy}, {Marois}, {Mennesson}, \&
  {Pueyo}}]{Mawet2014}
{Mawet}, D., {Milli}, J., {Wahhaj}, Z., {et~al.} 2014, ApJ, 792, 97

\bibitem[{{Mesa} {et~al.}(2018){Mesa}, {Baudino}, {Charnay}, {D'Orazi},
  {Desidera}, {Boccaletti}, {Gratton}, {Bonnefoy}, {Delorme}, {Langlois},
  {Vigan}, {Zurlo}, {Maire}, {Janson}, {Antichi}, {Baruffolo}, {Bruno},
  {Cascone}, {Chauvin}, {Claudi}, {De Caprio}, {Fantinel}, {Farisato}, {Feldt},
  {Giro}, {Hagelberg}, {Incorvaia}, {Lagadec}, {Lagrange}, {Lazzoni}, {Lessio},
  {Salasnich}, {Scuderi}, {Sissa}, \& {Turatto}}]{Mesa2018}
{Mesa}, D., {Baudino}, J.-L., {Charnay}, B., {et~al.} 2018, \aap, 612, A92

\bibitem[{{Mesa} {et~al.}(2015){Mesa}, {Gratton}, {Zurlo}, {Vigan}, {Claudi},
  {Alberi}, {Antichi}, {Baruffolo}, {Beuzit}, {Boccaletti}, {Bonnefoy},
  {Costille}, {Desidera}, {Dohlen}, {Fantinel}, {Feldt}, {Fusco}, {Giro},
  {Henning}, {Kasper}, {Langlois}, {Maire}, {Martinez}, {Moeller-Nilsson},
  {Mouillet}, {Moutou}, {Pavlov}, {Puget}, {Salasnich}, {Sauvage}, {Sissa},
  {Turatto}, {Udry}, {Vakili}, {Waters}, \& {Wildi}}]{Mesa2015}
{Mesa}, D., {Gratton}, R., {Zurlo}, A., {et~al.} 2015, A\&A, 576, A121

\bibitem[{{Milli} {et~al.}(2017){Milli}, {Hibon}, {Christiaens}, {Choquet},
  {Bonnefoy}, {Kennedy}, {Wyatt}, {Absil}, {G{\'o}mez Gonz{\'a}lez}, {del
  Burgo}, {Matr{\`a}}, {Augereau}, {Boccaletti}, {Delacroix}, {Ertel}, {Dent},
  {Forsberg}, {Fusco}, {Girard}, {Habraken}, {Huby}, {Karlsson}, {Lagrange},
  {Mawet}, {Mouillet}, {Perrin}, {Pinte}, {Pueyo}, {Reyes}, {Soummer},
  {Surdej}, {Tarricq}, \& {Wahhaj}}]{Milli2017a}
{Milli}, J., {Hibon}, P., {Christiaens}, V., {et~al.} 2017, \aap, 597, L2

\bibitem[{{Mizuno}(1980)}]{Mizuno1980}
{Mizuno}, H. 1980, Progress of Theoretical Physics, 64, 544

\bibitem[{{Mo{\'o}r} {et~al.}(2006){Mo{\'o}r}, {{\'A}brah{\'a}m}, {Derekas},
  {Kiss}, {Kiss}, {Apai}, {Grady}, \& {Henning}}]{Moor2006}
{Mo{\'o}r}, A., {{\'A}brah{\'a}m}, P., {Derekas}, A., {et~al.} 2006, \apj, 644,
  525

\bibitem[{{Mo{\'o}r} {et~al.}(2015){Mo{\'o}r}, {K{\'o}sp{\'a}l},
  {{\'A}brah{\'a}m}, {Apai}, {Balog}, {Grady}, {Henning}, {Juh{\'a}sz}, {Kiss},
  {Krivov}, {Pawellek}, \& {Szab{\'o}}}]{Moor2015}
{Mo{\'o}r}, A., {K{\'o}sp{\'a}l}, {\'A}., {{\'A}brah{\'a}m}, P., {et~al.} 2015,
  MNRAS, 447, 577

\bibitem[{{Mo{\'o}r} {et~al.}(2011){Mo{\'o}r}, {Pascucci}, {K{\'o}sp{\'a}l},
  {{\'A}brah{\'a}m}, {Csengeri}, {Kiss}, {Apai}, {Grady}, {Henning}, {Kiss},
  {Bayliss}, {Juh{\'a}sz}, {Kov{\'a}cs}, \& {Szalai}}]{Moor2011}
{Mo{\'o}r}, A., {Pascucci}, I., {K{\'o}sp{\'a}l}, {\'A}., {et~al.} 2011, \apjs,
  193, 4

\bibitem[{{Mordasini} {et~al.}(2012){Mordasini}, {Alibert}, {Benz}, {Klahr}, \&
  {Henning}}]{Mordasini2012}
{Mordasini}, C., {Alibert}, Y., {Benz}, W., {Klahr}, H., \& {Henning}, T. 2012,
  \aap, 541, A97

\bibitem[{{Mordasini} {et~al.}(2010){Mordasini}, {Klahr}, {Alibert}, {Benz}, \&
  {Dittkrist}}]{Mordasini2010}
{Mordasini}, C., {Klahr}, H., {Alibert}, Y., {Benz}, W., \& {Dittkrist}, K.-M.
  2010, ArXiv e-prints [\eprint[arXiv]{1012.5281}]

\bibitem[{{Mugnier} {et~al.}(2009){Mugnier}, {Cornia}, {Sauvage}, {Rousset},
  {Fusco}, \& {V{\'e}drenne}}]{Mugnier2009}
{Mugnier}, L.~M., {Cornia}, A., {Sauvage}, J.-F., {et~al.} 2009, Journal of the
  Optical Society of America A, 26, 1326

\bibitem[{{Mustill} \& {Wyatt}(2012)}]{Mustill2012}
{Mustill}, A.~J. \& {Wyatt}, M.~C. 2012, MNRAS, 419, 3074

\bibitem[{{Pace}(2013)}]{Pace2013}
{Pace}, G. 2013, \aap, 551, L8

\bibitem[{{Papaloizou} {et~al.}(2001){Papaloizou}, {Nelson}, \&
  {Masset}}]{Papaloizou2001}
{Papaloizou}, J.~C.~B., {Nelson}, R.~P., \& {Masset}, F. 2001, \aap, 366, 263

\bibitem[{{Pavlov} {et~al.}(2008){Pavlov}, {M{\"o}ller-Nilsson}, {Feldt},
  {Henning}, {Beuzit}, \& {Mouillet}}]{Pavlov2008}
{Pavlov}, A., {M{\"o}ller-Nilsson}, O., {Feldt}, M., {et~al.} 2008, in SPIE
  Conf. Ser., Vol. 7019, 701939

\bibitem[{{Pawellek} \& {Krivov}(2015)}]{Pawellek2015}
{Pawellek}, N. \& {Krivov}, A.~V. 2015, \mnras, 454, 3207

\bibitem[{{Pawellek} {et~al.}(2014){Pawellek}, {Krivov}, {Marshall},
  {Montesinos}, {{\'A}brah{\'a}m}, {Mo{\'o}r}, {Bryden}, \&
  {Eiroa}}]{Pawellek2014}
{Pawellek}, N., {Krivov}, A.~V., {Marshall}, J.~P., {et~al.} 2014, \apj, 792,
  65

\bibitem[{{Pearce} {et~al.}(2014){Pearce}, {Wyatt}, \& {Kennedy}}]{Pearce2014}
{Pearce}, T.~D., {Wyatt}, M.~C., \& {Kennedy}, G.~M. 2014, MNRAS, 437, 2686

\bibitem[{{Pearce} {et~al.}(2015){Pearce}, {Wyatt}, \& {Kennedy}}]{Pearce2015}
{Pearce}, T.~D., {Wyatt}, M.~C., \& {Kennedy}, G.~M. 2015, \mnras, 448, 3679

\bibitem[{{Petrovich}(2015)}]{Petrovich2015}
{Petrovich}, C. 2015, \apj, 808, 120

\bibitem[{{Pollack} {et~al.}(1996){Pollack}, {Hubickyj}, {Bodenheimer},
  {Lissauer}, {Podolak}, \& {Greenzweig}}]{Pollack1996}
{Pollack}, J.~B., {Hubickyj}, O., {Bodenheimer}, P., {et~al.} 1996, Icarus,
  124, 62

\bibitem[{{Rameau} {et~al.}(2013){Rameau}, {Chauvin}, {Lagrange}, {Meshkat},
  {Boccaletti}, {Quanz}, {Currie}, {Mawet}, {Girard}, {Bonnefoy}, \&
  {Kenworthy}}]{Rameau2013c}
{Rameau}, J., {Chauvin}, G., {Lagrange}, A.-M., {et~al.} 2013, \apjl, 779, L26

\bibitem[{{Reg{\'a}ly} {et~al.}(2018){Reg{\'a}ly}, {Dencs}, {Mo{\'o}r}, \&
  {Kov{\'a}cs}}]{Regaly2018}
{Reg{\'a}ly}, Z., {Dencs}, Z., {Mo{\'o}r}, A., \& {Kov{\'a}cs}, T. 2018,
  \mnras, 473, 3547

\bibitem[{{Rhee} {et~al.}(2007){Rhee}, {Song}, {Zuckerman}, \&
  {McElwain}}]{Rhee2007}
{Rhee}, J.~H., {Song}, I., {Zuckerman}, B., \& {McElwain}, M. 2007, ApJ, 660,
  1556

\bibitem[{{Th{\'e}bault} \& {Wu}(2008)}]{Thebault2008}
{Th{\'e}bault}, P. \& {Wu}, Y. 2008, \aap, 481, 713

\bibitem[{{Toomre}(1964)}]{Toomre1964}
{Toomre}, A. 1964, \apj, 139, 1217

\bibitem[{{Vigan} {et~al.}(2010){Vigan}, {Moutou}, {Langlois}, {Allard},
  {Boccaletti}, {Carbillet}, {Mouillet}, \& {Smith}}]{Vigan2010}
{Vigan}, A., {Moutou}, C., {Langlois}, M., {et~al.} 2010, MNRAS, 407, 71

\bibitem[{{Wisdom}(1980)}]{Wisdom1980}
{Wisdom}, J. 1980, \aj, 85, 1122

\bibitem[{Wyatt(2005)}]{Wyatt2005}
Wyatt, M.~C. 2005, Astronomy \& Astrophysics, 440, 937

\end{thebibliography}
%

\begin{appendix}

\section{Dynamical simulation results for a slightly non-coplanar companion}
\label{sec:dynamics_noncoplanar}

\begin{figure}[h]
\centering
\includegraphics[width=.49\linewidth]{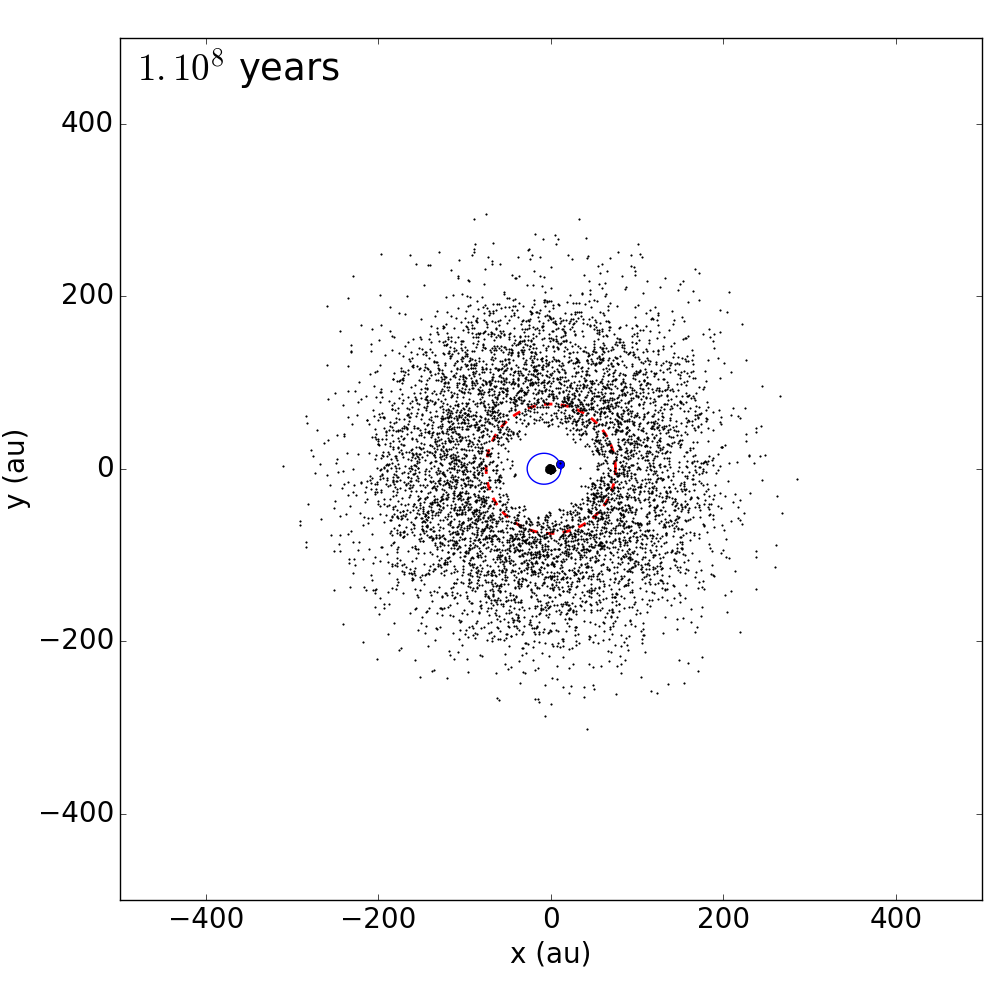}
\includegraphics[width=.49\linewidth]{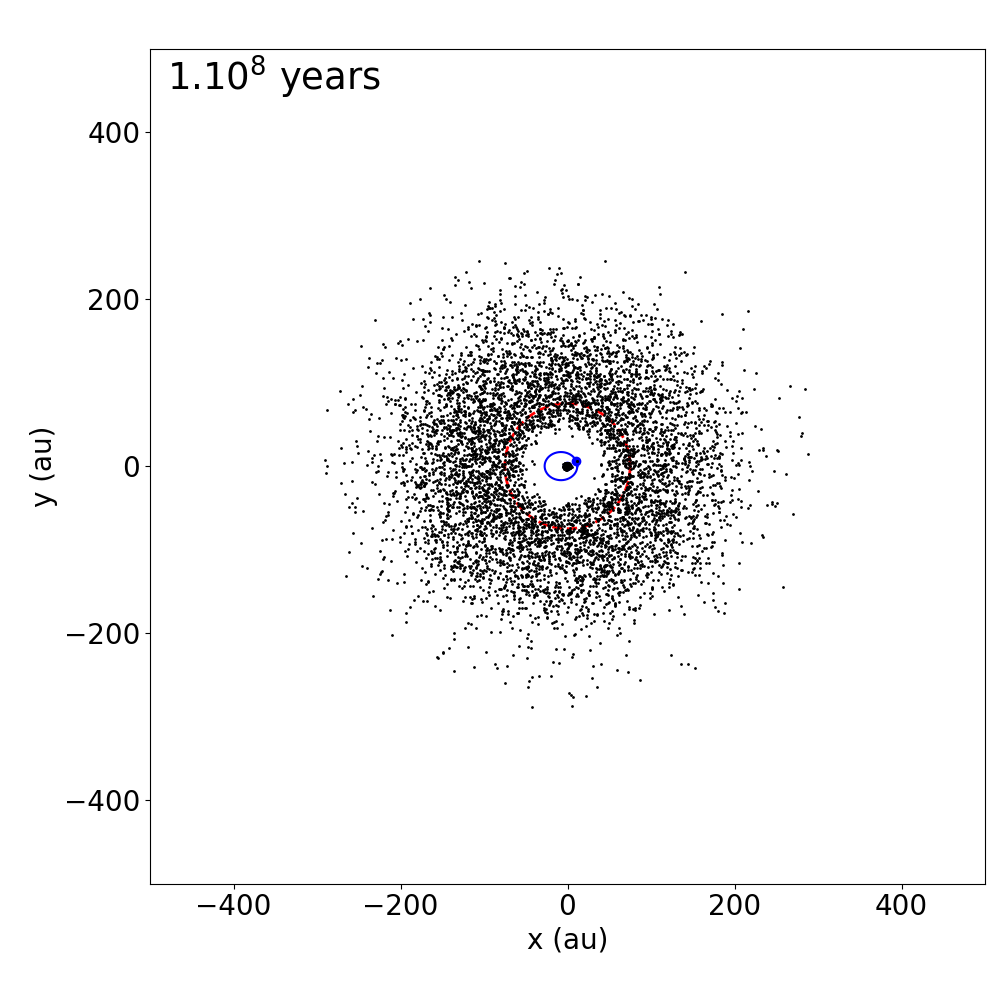}
\caption{Simulated images of the disk for an eccentric ($e$=0.4) companion orbit of semi-major axis 20~au coplanar with the disk (\textit{left}) and with a relative inclination of 20$^{\circ}$ (\textit{right}).}
\label{fig:dynamics_noncoplanar}
\end{figure}

Figure~\ref{fig:dynamics_noncoplanar} shows the simulated images of the disk for a companion orbit with a semi-major axis of 20~au and an eccentricity of 0.4 for a coplanar configuration and a relative inclination of 20$^{\circ}$ with the disk. The simulated relative inclination has no significant effects on the cavity size. The only difference is that in the non-coplanar case, the companion stirs the inclination of the disk particles, which gives the disk a non-negligible thickness (which cannot be seen on the face-on image).

\section{Width of the chaotic zone predicted by other relations}
\label{sec:petrovich2015regaly2018}

\begin{figure*}[t]
\centering
\includegraphics[trim=10mm 0mm 8mm 10mmm,width=0.4\textwidth]{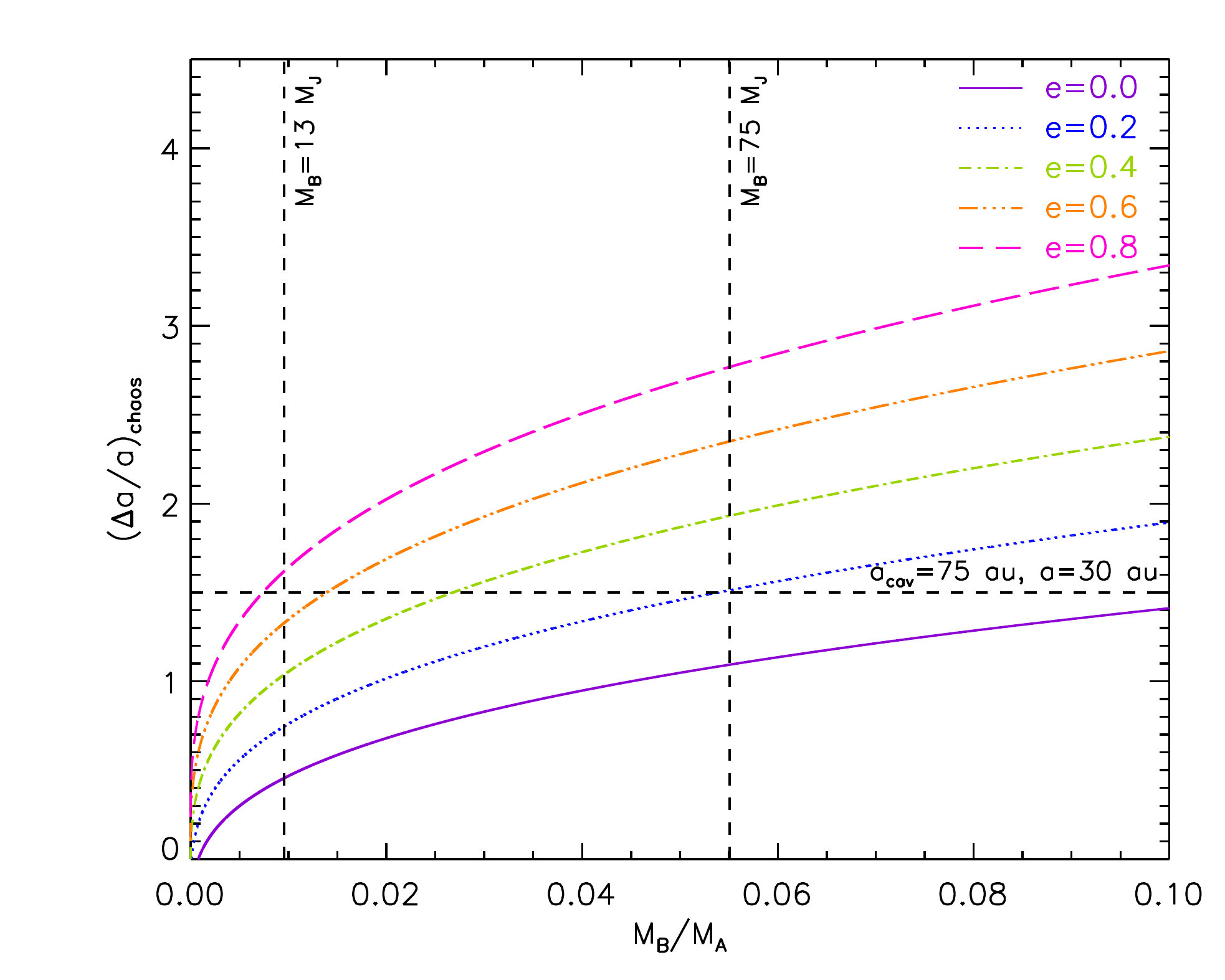}
\includegraphics[trim=10mm 0mm 8mm 10mmm,width=0.4\textwidth]{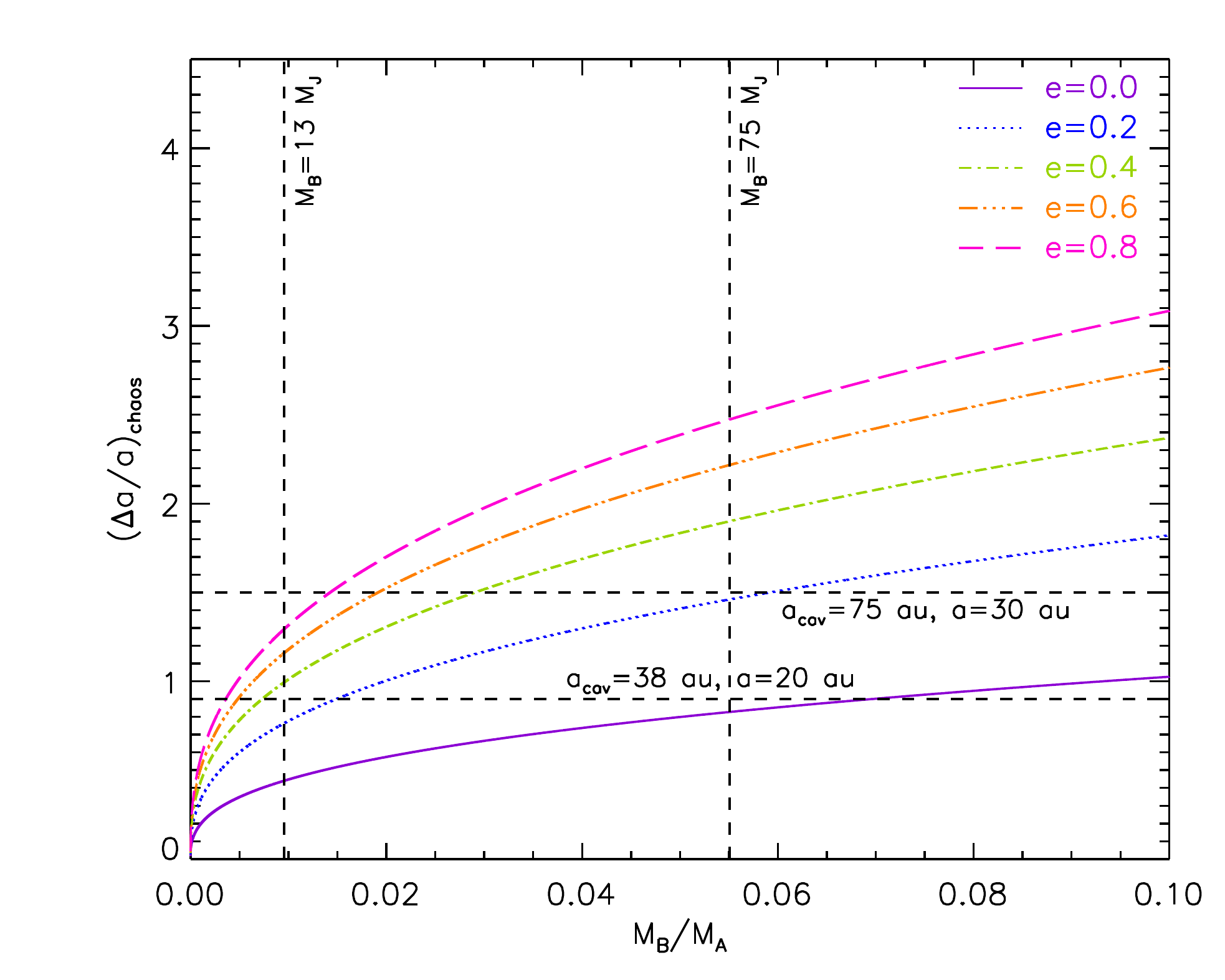}
\caption{{Same as Fig.~\ref{fig:wisdommustill} but for the relations in \citet{Petrovich2015} (\textit{left}) and \citet{Regaly2018} (\textit{right}).}}
\label{fig:petrovich2015regaly2018}
\end{figure*}

{We show in Fig.~\ref{fig:petrovich2015regaly2018} the width of the chaotic zone created by a substellar companion in the disk of HR~2562 for several eccentricities according to the empirical relations of \citet{Petrovich2015} and \citet{Regaly2018}.}

{We first used the equation in \citet{Petrovich2015}, which is an empirical dynamical stability criterion for two-planet systems against collisions with the star and/or ejections from the system. The formula was validated using numerical simulations for planet/star mass ratios 10$^{-4}$--10$^{-2}$ and mutual inclinations $\lesssim$40$^{\circ}$. In order to apply this relation to the HR~2562 companion-disk system, we assumed that the outer planet has a negligible mass \citep[the most extreme mass ratio between the planets is 1/100 in][]{Petrovich2015}. {The 1.15 constant term in the formula includes a margin of 0.5 to account for disk regions which are potentially unstable. In order not to overestimate the cavity size and to make the comparison to the criteria in \citet{Lazzoni2018} and \citet{Regaly2018} coherent, we therefore decreased the constant term in the formula of \citet{Petrovich2015} by 0.5. We finally assumed} a null eccentricity for the debris belt, which is exterior to the companion. Contrary to the formula in \citet{Lazzoni2018} and \citet{Regaly2018}, the formula of \citet{Petrovich2015} depends on the cavity radius and companion semi-major axis. We chose $a_{\rm{cav}}$\,=\,75~au and $a$\,=\,30~au. We represent the resulting curves in the left panel of Fig.~\ref{fig:petrovich2015regaly2018}. The eccentricity is not well constrained with respect to the predictions of \citet{Lazzoni2018} (Fig.~\ref{fig:wisdommustill}) and can lie in the range $\sim$0.2--0.7.}

{Subsequently, we considered the relations in the recent work of \citet{Regaly2018}, which predict the size of the cavity of a debris disk shaped by a giant planet perturber interior to the debris belt. They were determined using N-body simulations assuming a giant planet with mass ratios to the star 1.25$\times$10$^{-3}$--10$^{-2}$ with eccentricities 0--0.9. Quasi-circular orbits cannot be excluded for a disk cavity size of 38~au and a companion semi-major axis of 20~au, whereas eccentricities as large as $\sim$0.3 are allowed for the smallest mass range compatible with a brown dwarf. For a disk cavity size of 75~au and a companion semi-major axis of 30~au, the eccentricity is also poorly constrained with respect to the predictions of \citet{Lazzoni2018} and can range from $\sim$0.2 up to more than 0.8.}

{The more stringent constraints on the companion eccentricity obtained using the relations in \citet{Lazzoni2018} stem from the flatter global slopes of the relations with respect to those in \citet{Petrovich2015} and \citet{Regaly2018}. The relations in \citet{Lazzoni2018} predict wider chaotic zones at mass ratios below $\sim$0.02 and eccentricities larger than 0.2 with respect to the equations of \citet{Petrovich2015} and \citet{Regaly2018} while predict narrower chaotic zones for mass ratios larger than $\sim$0.04 and eccentricities smaller than 0.6.}

{We finally note that the relations in \citet{Petrovich2015} and \citet{Regaly2018} usually predict similar values for the chaotic zone widths, except for a circular orbit and large companion/star mass ratios ($\gtrsim$0.03) and for highly-eccentric orbits ($\gtrsim$0.6).}

\end{appendix}

\end{document}